\documentclass[superscriptaddress,nofootinbib,preprintnumbers,floatfix,twocolumn,prd]{revtex4}

\usepackage{hyperref}
\usepackage{bm}
\usepackage{amsmath}
\usepackage{graphicx}
\usepackage{psfrag}
\usepackage[usenames]{xcolor}
\usepackage{subfigure}
\usepackage[toc,page]{appendix}

\newcommand{\eq}[1]{Eq.~\eqref{eq:#1}}
\newcommand{\eqs}[2]{Eqs.~\eqref{eq:#1} and \eqref{eq:#2}}
\renewcommand{\sec}[1]{Sec.~\ref{sec:#1}}

\newcommand{\subsec}[1]{Sec.~\ref{subsec:#1}}
\newcommand{\fig}[1]{Fig.~\ref{fig:#1}}
\newcommand{\figs}[2]{Figs.~\ref{fig:#1} and \ref{fig:#2}}

\newcommand{\sdt}{\!\cdot\!}
\newcommand{\cP}{{\mathcal P}}

\newcommand{\bfd}{{\bm d}}
\newcommand{\bfk}{{\bm k}}
\newcommand{\bfl}{{\bm \ell}}
\newcommand{\bfL}{{\bm L}}
\newcommand{\bfm}{{\bm m}}

\newcommand{\bfx}{{\bm x}}
\newcommand{\ord}[1]{{\mathcal O}(#1)}
\newcommand{\intlnp}{\int\! \frac{d^2 {\bfl^\prime}}{(2\pi)^2}\,}
\newcommand{\intlnpp}{\int\! \frac{d^2 {\bfl^{\prime\prime}}}{(2\pi)^2}\,}

\newcommand{\ff}[3]{f^{(#2)#1}_{#3}}

\newcommand{\zero}{{(0)}}
\newcommand{\one}{{(1)}}

\newcommand{\intl}{\int\! \frac{d^2 \bfl}{(2\pi)^2}}
\newcommand{\intk}{\int\! \frac{d^2 \bfk}{(2\pi)^2}}
\newcommand{\intm}{\int\! \frac{d^2 \bfm}{(2\pi)^2}}

\newcommand{\threej}[6]{ \left( \begin{array}{ccc}
#1 & #3 & #5 \\
#2 & #4 & #6
\end{array} \right)}

\def\nhat{ \hat {\bf n}}

\def\nn{\nonumber\\ }
\def\rd{{\rm d}}

\def\vev#1{\left\langle #1 \right \rangle}

\def\corr#1#2#3#4{{\rm #1#2\hspace{0.15em}#3#4}}
\def\est#1#2{{\rm #1#2}}

\def\planck{{\sc Planck}}

\begin{document}


\title{Higher-Order Gravitational Lensing Reconstruction using Feynman Diagrams}

\preprint{\vbox{
\hbox{NIKHEF 2014-007}}}

\author{Elizabeth E.~Jenkins}
\affiliation{Department of Physics, University of California at San Diego, La Jolla, CA 92093, USA}

\author{Aneesh V.~Manohar}
\affiliation{Department of Physics, University of California at San Diego, La Jolla, CA 92093, USA}

\author{Wouter J.~Waalewijn}
\affiliation{Nikhef, Theory Group, Science Park 105, 1098 XG, Amsterdam, The Netherlands}
\affiliation{ITFA, University of Amsterdam, Science Park 904, 1018 XE, Amsterdam, The Netherlands}

\author{Amit P.~S.~Yadav}
\thanks{\mbox{Corresponding author. \href{mailto:ayadav@physics.ucsd.edu}{ayadav@physics.ucsd.edu}}}
\affiliation{Department of Physics, University of California at San Diego, La Jolla, CA 92093, USA}

\date{\today}

\begin{abstract}

We develop a method for calculating the correlation structure of the Cosmic Microwave Background (CMB) using Feynman diagrams, when the CMB has been modified by gravitational lensing, Faraday rotation, patchy reionization, or other distorting effects. This method is used to calculate the bias of the Hu-Okamoto quadratic estimator in reconstructing the lensing power spectrum up to $\ord{\phi^4}$ in the lensing potential $\phi$. We consider both the diagonal noise \corr T T T T, \corr E B E B, etc.~and, for the first time, the off-diagonal noise \corr T T  T E,  \corr T B E B, etc. The previously noted large $\ord{\phi^4}$ term in the second order noise is identified to come from a particular class of diagrams. It can be significantly reduced by a reorganization of the $\phi$ expansion. These improved estimators have almost no bias for the off-diagonal case involving only one $B$ component of the CMB, such as \corr E E E B.

\end{abstract}

\maketitle

\section{Introduction}\label{sec:intro}

Primary anisotropies in the Cosmic Microwave Background (CMB) were generated around $375,000$ years after the ``big-bang", when the perturbations were still in the linear regime. The CMB is characterized by its temperature and polarization. The temperature map $T(\nhat)$ describes the temperature fluctuation in the direction $\nhat$. The polarization of the CMB radiation is conventionally decomposed in terms of the  polarization modes $E$ and $B$, which have even and odd parity, respectively. Primordial scalar perturbations create only $E$-modes of the CMB, while primordial tensor perturbations generate both parity-even $E$-modes and parity-odd $B$-modes~\cite{Seljak:1996gy, Kamionkowski:1996ks, Kamionkowski:1996zd}. The amplitude of primordial $B$-modes of the CMB is proportional to the energy scale at which inflation occurs; hence constraints on $B$-modes will provide valuable information about the early universe~\cite{2009AIPC.1141...10B}. 

Many ground-based and balloon experiments are looking for primordial $B$-modes of the CMB. 
The primordial $B$-mode gravitational wave signal has just been detected, constraining the ratio of tensor to scalar perturbations~\cite{Ade:aa}. 
Even in the absence of primordial $B$-modes, subsequent gravitational lensing by the Large Scale Structure (LSS) of the Universe converts $E$-mode polarization to $B$-mode polarization~\cite{1998PhRvD..58b3003Z}. Gravitational lensing of the CMB was first detected by cross-correlating the CMB with LSS data~\cite{2007PhRvD..76d3510S, 2008PhRvD..78d3520H}. It has since been detected using CMB data alone~\cite{2011PhRvL.107b1301D,Ade:2013tyw,2013PhRvL.111n1301H, 2013arXiv1312.6646P, 2013arXiv1312.6645P}.

CMB lensing is a clean and powerful probe of several cosmological parameters~\cite{1997MNRAS.291L..33B,1997ApJ...488....1Z,1998ApJ...492L...1M,Kaplinghat:2003bh, 2006PhRvD..74j3510A,2002PhRvD..65b3003H}, see e.g.~the reviews in Refs.~\cite{Lewis:2006fu,2009AIPC.1141..121S}. Lensing depends on the integrated (time-dependent) gravitational potential along the path of the CMB photons, and
provides the deepest possible measurement of the two dimensional mass distribution of the universe. It can be used to break degeneracies between cosmological parameters and to constrain the neutrino mass, independently of other probes like Lyman-$\alpha$ and galaxy clustering. It also probes the late time evolution of  structure in the universe, and hence provides constraints on dark energy and the integrated Sachs-Wolfe effect. Finally, if the lensing map is well measured, then one can de-lens the observed CMB (i.e.~effectively subtract off lensing induced $B$-modes) to increase sensitivity to inflationary $B$-modes and the energy scale of inflation.

Initial studies focused on the effect of lensing on the power spectra of the CMB modes~\cite{Seljak:1995ve,Zaldarriaga:1998ar,Seljak:1998nu}.  Later, the emphasis switched to constraining the LSS from measurements of the CMB, by using e.g.~the fact that lensing introduces non-Gaussian fluctuations~\cite{Zaldarriaga:2000ud} which are not present in the primordial CMB. A particularly fruitful approach has been the use of quadratic estimators, built out of the convolution of two CMB modes to determine the lensing from the statistical breaking of isotropy. A quadratic estimator based on CMB modes in position space was introduced in Ref.~\cite{Seljak:1998aq}. We  focus on the Hu-Okamoto quadratic estimator~\cite{Hu:2001kj,Okamoto:2003zw}, which is uniquely determined by the requirements that it is unbiased and has minimal variance. A likelihood-based approach shows that for the temperature mode this method is close to optimal, though improvement is possible for polarization modes~\cite{Hirata:2002jy,Hirata:2003ka}. The quadratic estimator has been used by \planck\ to detect lensing at $>25\,\sigma$ using only temperature information~\cite{Ade:2013tyw}. Recently the detection of lensing using  CMB polarization data has been reported in Refs.~\cite{2013PhRvL.111n1301H, 2013arXiv1312.6646P, 2013arXiv1312.6645P}.

The quadratic estimators provide an unbiased means for extracting the lensing potential $\phi$. However, an estimate of the lensing power spectrum depends on the two-point function of quadratic estimators. This two-point function will have a bias due to inherent noise from random fluctuations. This noise bias needs to be subtracted to get a reliable measurement of the lensing power spectrum.
The calculation of noise is carried out in a small lensing expansion (unlike the simpler case of the lensed power spectra where such an expansion is unnecessary~\cite{Challinor:2005jy}). The first order correction for all estimators, and the second order correction for the estimator \est T T\ were determined in Refs.~\cite{Cooray:2002py,Kesden:2003cc,Hanson:2010rp,2014arXiv1403.2386J}. Simulations of the second order corrections for all channels were studied in Ref.~\cite{Anderes:2013jw}. The second-order corrections turned out to be unexpectedly large at small $L$ (large scales), casting doubt on the convergence of the small lensing expansion.

In this paper, we develop the method for calculating the lensing of the CMB using Feynman diagrams presented in Ref.~\cite{2014arXiv1403.2386J}, which simplifies its calculation and enables us to readily identify the origin of this large second-order correction. Feynman diagrams have been used in other cosmological applications, see e.g.~\cite{Goroff:1986ep}, and were employed in Ref.~\cite{Rathaus:2011xi} to study the effect of a single gravitational lens.
The computation of CMB correlations is very similar to the computation of correlation functions in statistical physics or quantum field theory, so the standard Feynman diagram approach is a very convenient and efficient way to organize the calculation. The lensing potential can be extracted using the $xy$ quadratic estimator, where $x,y \in \{T,E,B\}$, and the lensing power spectrum can be extracted by looking at the $xy-x^\prime y^\prime$ two-point function. Our formalism can be applied to study the noise not only for the \corr T T T T\ correlator, but for all of the $xy-x^\prime y^\prime$ correlators. 

In addition to gravitational lensing, the observed CMB can be distorted by other effects, and our diagrammatic treatment can be trivially extended to these cases. We will discuss two cases in \sec{other}, focussing on distortions due to patchy reionization, and rotations of the CMB polarization due to parity-violating Chern-Simons terms from axion fields, which exist in many theories.

The outline of this paper is as follows. In \sec{lensing}, we review the basics of gravitational lensing of the CMB, which we formulate in the language of Feynman diagrams in \sec{feynman}. We derive the Feynman rules for patchy reionization and rotation of the CMB polarization in \sec{other}. The quadratic estimator for gravitational lensing and its  noise terms up to order $\phi^4$ are calculated in \sec{quadratic}. We show numerical results in \sec{results} and conclude in \sec{summary}.

\section{Gravitational Lensing of the CMB}
\label{sec:lensing}

In this section, we review the basic lensing calculation for both temperature and
polarization fields. 
More details of CMB lensing can be found in the excellent
reviews of Refs.~\cite{Lewis:2006fu,2009AIPC.1141..121S}.

Gravitational lensing deflects the path of CMB photons from the last 
scattering surface at $z \sim 1090$ to us, resulting in a remapping of the CMB
temperature and polarization pattern on the sky.
We will use a tilde to denote the observed (lensed) fields, which at a certain position $\nhat$ on the sky
are related to the (unlensed) primordial CMB fields by
\begin{align} \label{eq:RealTay}
\widetilde T(\nhat) & = T\big(\nhat + \bfd(\nhat)\big) 
\,,\nn
\widetilde Q(\nhat) & = Q\big(\nhat +\bfd(\nhat)\big)
\,,\nn
\widetilde U(\nhat) & = U\big(\nhat +\bfd(\nhat)\big)
\,.\end{align}
Here $T$ is the temperature fluctuation, $Q$ and $U$ are the Stokes parameters encoding the polarization, and $\bfd(\nhat)$ is the deflection. The average deflection angle, $\langle \bfd\cdot \bfd\rangle^{1/2}$, is of the order of a few arc-minutes. However, the deflection is coherent over much larger scales, with power peaking at the scale of a few degrees. The lensing remapping process conserves the surface brightness distribution of the CMB, and thus does not change the one-point statistics. 
 
The deflection angle is related to the lensing gravitational potential
$\phi(\nhat)$ by
\begin{equation} \label{eq:deflection}
 \bf{d}(\nhat) = \nabla\phi(\nhat)
\,, \end{equation}
where the gradient $\nabla$ is with respect to $\nhat$. 
In the small lensing approximation, the
lensing potential $\phi(\nhat)$ is given by
an integral of the zero-shear gravitational potential $\Phi$ along the line of sight,
\begin{align} \label{eq:lenspotential}
 \phi(\nhat)&= - 2 \int_0^{r_0}\! dr\,
 \frac{d_A(r_0-r)}{d_A(r)d_A(r_0)} \Phi (-r, r  \nhat)
\,. \end{align}
Here $r(z)$ is the comoving distance at redshift $z$
\begin{align}
r(z)=\int^z_0\! \frac{dz'}{H(z')}
\,,\end{align}
$r_0$ is the position of the last scattering surface, and
$d_A$ is the comoving angular diameter distance,
\begin{align}
d_A(r)=H^{-1}_0 \Omega^{-1/2}_K \text{sinh}(H_0\Omega^{1/2}_K r)
\,,\end{align}
in units where the speed of light $c=1$.
In the limit $\Omega_K \to 0$ that the curvature vanishes, $d_A(r) \to r$. 
Beyond lowest order in the lensing, there are corrections to the lensing potential in \eq{lenspotential}~\cite{Bernardeau:1996un,Cooray:2002mj} and additional curl-type (or shear) contributions to the deflection in \eq{deflection}~\cite{Hirata:2003ka,Cooray:2005hm,Namikawa:2011cs}.
We note that the  lensing potential $\phi$ can be related to the convergence $\kappa$ of the light rays by
\begin{align}
\kappa(\nhat) &= -\frac12\, \nabla^2\phi(\nhat) 
\,, \end{align}
which is typically used in studies involving the galaxy shear.

We formulate CMB lensing using the flat-sky
approximation. The flat-sky approach simplifies the
derivation by replacing summations over Wigner symbols of spherical
harmonics by integrals involving mode coupling
angles~\cite{Hu:2000ee}. The basic difference arises from integrals
\begin{align}
& \int \rd \Omega\ Y_{l_1 m_1}(\Omega) Y_{l_2 m_2}(\Omega) Y_{l_3 m_3}(\Omega) \nn
 &= 
\sqrt{ \frac{(2l_1+1)(2l_2+1)(2l_3+1)}{4\pi}} \nn
&\qquad \times \threej {l_1}{m_1}{l_2}{m_2}{l_3}{m_3} \threej {l_1}{0}{l_2}{0}{l_3}{0} 
\label{eq:full}
\end{align}
for the full-sky, which simplify, in the flat-sky approximation to
\begin{align} \label{eq:flat}
\int  \rd^2 \mathbf{x} \ e^{i\left( \bfl_1 + \bfl_2 + \bfl_3 \right) \cdot \mathbf{x} }
&= (2\pi)^2 \delta^{(2)} ( \bfl_1 + \bfl_2 + \bfl_3)
\,.\end{align}
Here $\mathbf{x}$ is the point in a plane that corresponds to $\nhat$  by a stereographic projection. In the flat sky approximation, the power spectrum in the momentum $\bfk$ conjugate to $\mathbf{x}$ is equal to that of the spherical harmonic with multipole moment $L = |\bfk|$.
Eq.~(\ref{eq:full}) is the quantum mechanical form for the conservation of angular-momentum, which reduces in the flat-sky approximation to \eq{flat}, which is analogous to the conservation of linear momentum, with angular momentum $\bfl$ assuming the role of linear momentum.  One can derive expressions similar to \eq{full} for the integral of more than three spherical harmonics in terms of Wigner $nj$ symbols.

An important advantage of Feynman diagrams is the simplification of combinatorics in evaluating  correlation functions. The combinatorics of the diagrammatic expansion is the same for the flat-sky approximation and for the full-sky calculation, so the graphical representation is unchanged. The only difference between the flat-sky and full-sky computations is the use of \eq{full}  instead of \eq{flat} for angular momentum conservation at the vertices of diagrams.

It is convenient to work in Fourier (i.e.~angular momentum) space. In the flat-sky approximation, the Fourier modes become plane waves, and
\begin{align} \label{eq:fourier}
 T_\bfl
&=\int\! d\bfx \, T(\bfx)\, e^{-i \bfl \cdot \bfx} \nn
 E_\bfl \pm i  B_\bfl  &=
 \int\! d \bfx \, [ Q(\bfx)\pm i  U(\bfx)]\, e^{\mp 2i\varphi_{\bfl}} e^{-i \bfl \cdot \bfx}\,,\nn
\phi_\bfl &=\int d \bfx\,  \phi(\bfx)\, e^{-i \bfl \cdot \bfx}\,.
\end{align}
Here, $\bfl$ denotes the Fourier mode and $\varphi_{\bfl}=\cos^{-1}(\hat x \cdot \bfl)$. 
In the second equation, we have related the Fourier transform of the polarization modes $E_\bfl$ and $B_\bfl$ to the Stokes parameters $Q$ and $U$ in position space.  The
factor $ e^{\mp 2i\varphi_{\bf \ell}}$ enters in the Fourier transform definition
since $\left[ Q\pm iU \right]$ is a spin-2 field.

Taylor expanding~\eq{RealTay} yields
\begin{widetext}
\begin{align}\label{eq:lensD}
\widetilde T_\bfl
&= T_\bfl + \intlnp T_{\bfl'} \bigg\{- \phi_{\bfl-\bfl'} \, \left[(\bfl-\bfl') \cdot \bfl' \right] 
+\frac{1}{2} \intlnpp \phi_{\bfl''} \phi_{\bfl - \bfl' - \bfl''}  (\bfl'' \cdot
\bfl') \left[ (\bfl - \bfl' - \bfl'')\cdot \bfl' \right] + \ldots \bigg\} 
\,,\nn
[\widetilde E_{\bfl} \pm i \widetilde B_{\bfl}] &= [E_{\bfl} \pm i B_{\bfl}] + \intlnp [E_{\bfl'} \pm i B_{\bfl'}] \bigg\{ -e^{\pm2i(\varphi_{\bfl'}-\varphi_\bfl)}\phi_{\bfl-\bfl'}  \left[
(\bfl-\bfl') \cdot \bfl' \right] 
\nn & \quad
+ \frac{1}{2} \intlnpp
e^{\pm2i(\varphi_{\bfl'}-\varphi_\bfl)} \phi_{\bfl''} \phi_{\bfl - \bfl' - \bfl''}  (\bfl'' \cdot
\bfl')\left[ (\bfl - \bfl' - \bfl'')\cdot \bfl' \right]+ \ldots  \bigg\} 
\,.\end{align}
\end{widetext}
The remapping of CMB fields induced by the gradient of the lensing potential leads to a convolution in Fourier space which couples different harmonic modes. The term with one $\phi$ field reproduces the familiar result of Ref.~\cite{Hu:2001kj}.

We now discuss the effect of gravitational lensing (and other distortions) on $n$-point functions.
The power spectrum, bispectrum, trispectrum and so on of the
CMB fields and the lensing potential are defined in the usual
manner by
\begin{align} \label{eq:cordef}
\big\langle x_{\bfl_1} y_{\bfl_2}\big\rangle &\equiv
        (2\pi)^2 \delta(\bfl_1+\bfl_2)  C^{xy}_{\bfl_1}\,,\nonumber\\
\big\langle x_{\bfl_1} y_{\bfl_2} z_{\bfl_3}\big\rangle_c &\equiv (2\pi)^2 \delta(\bfl_1+\bfl_2+\bfl_3)
        B^{xyz}_{\bfl_1,\bfl_2,\bfl_3}\,,\nonumber\\
\big\langle x_{\bfl_1} y_{\bfl_2} z_{\bfl_3} w_{\bfl_4}\big\rangle_c&\equiv (2\pi)^2 \delta(\bfl_1+\bfl_2+\bfl_3+\bfl_4)
        T^{xyzw}_{\bfl_1,\bfl_2,\bfl_3,\bfl_4}\,,\nonumber\\
 & \qquad       \ldots 
\end{align}
where
the angle brackets represent averages over realizations of any stochastic field under consideration --- primordial CMB, the LSS, or the experimental noise. We will use a subscript on the angle brackets if we want to make the averaging field explicit, e.g.~$\vev{\ }_{\text{CMB}}$, $\vev{\ }_{\text{LSS}}$, $\vev{\ }_{\text{CMB,LSS}}$ are averages over realizations of the CMB, LSS, and both CMB and LSS, respectively. The subscript $c$ denotes the connected part of the $n$-point function, and will often be dropped in the following discussion. The fields $x_\bfl$, $y_\bfl$, $z_\bfl$, $w_\bfl$ in \eq{cordef} can be any of $\phi_\bfl$, $T_\bfl$, $E_\bfl$, $B_\bfl$, $\widetilde T_\bfl$, $\widetilde E_\bfl$, $\widetilde B_\bfl$.

The primordial CMB is statistically isotropic and Gaussian, so all of the information is contained in the power spectrum $C^{xy}_\bfl$, 
which only depends on $\ell = |\bfl|$. The temperature and polarization components of the CMB, $x,y\in \{T,E,B\}$ can be combined conveniently into a column vector $X$ such that $C_{\bfl}^{xy}$ are components of a symmetric $3\times 3$ CMB power spectrum matrix $C_{\bfl}$,
\begin{align}
\langle X_\bfl\, X^T_\bfk \rangle_\text{CMB} &= \langle \begin{pmatrix}T_\bfl \\ E_\bfl  \\ B_\bfl\end{pmatrix} \begin{pmatrix}T_\bfk & E_\bfk &B_\bfk \end{pmatrix}\rangle_\text{CMB}
\nn 
&= C_\bfl (2\pi)^2 \delta^2(\bfl+\bfk)
\label{eq:cmbavg}
\,.\end{align}
Note that the off-diagonal $C^{TE}$ element of the matrix does not vanish, since there can be primordial \est T E\ correlations.
For a Gaussian primordial field, the bispectrum and higher order odd-correlations vanish, and the higher order even correlations can be written in terms of sums of products of the power spectrum. 

The presence of lensing or other distortions, such as patchy reionization and Faraday rotation, will convert the primordial Gaussian value $X$ into the components of $\widetilde X$~\cite{Yadav:2009za, 2009PhRvD..79d3003D, 2009PhRvD..79l3009Y,2009PhRvD..80b3510G, 2009PhRvL.102k1302K, 2012PhRvD..86l3009Y}, which is not Gaussian and which has a non-vanishing bispectrum, trispectrum, etc.  The CMB bispectrum~\cite{1999PhRvD..59j3002G,1999PhRvD..59j3001S, 1999PhRvD..59l3507Z} is generated because the gravitational potential correlates with other secondary effects such as the integrated Sachs-Wolfe effect~\cite{ISW} or the thermal Sunyaev-Zel'dovich effect~\cite{SZ}. Otherwise, it would vanish. The trispectrum, on the other hand, is generated by the non-linear nature of lensing itself~\cite{Zaldarriaga:2000ud}. The effect of distortion fields on Fourier modes of the CMB may be written generically as
\begin{equation}\label{eq:gen}
\widetilde X_\bfl = \intm\, D_{(\bfl,\bfm)}\,X_\bfm
\,,\end{equation}
where the $3 \times 3$ distortion matrix $D$ can have off-diagonal terms which mix components of the CMB. For example, $D^{BE}$ turns primordial $E$ modes into $B$ modes. For gravitational lensing, the deviation of $D$ from the identity matrix can be identified with the terms inside the curly brackets in~\eq{lensD}. The precise form of $D$ depends on the nature of the distorting field. The expression for $D$ due to gravitational lensing is given in \sec{feynman}.  In \sec{other}, we give $D$ when the distortions are due to patchy reionization or Faraday rotation.

In this paper, we will assume that the distorting field ${\cal D}$ is Gaussian and statistically isotropic such that it can be completely characterized by its power spectrum $C^{\cal D \cal D}_\bfl$. For lensing, the distorting field is $\bfd$, or equivalently, the lensing potential $\phi$, so $C^{\cal D \cal D}_\bfl$ is $C^{dd}_\bfl= \bfl^2 C^{\phi \phi}_\bfl$, where $C^{\phi \phi}_\bfl$ is the power spectrum of the gravitational potential,
\begin{align}\label{eq:lssavg}
\langle \phi_\bfl\, \phi_\bfk \rangle_\text{LSS} &= C_\bfl^{\phi \phi} (2\pi)^2 \delta^2(\bfl+\bfk)
\,.\end{align}
We consider only Gaussian distorting fields because they are the most important case of experimental interest. However, the formalism developed in this paper is not limited to this case, and it can be extended easily to non-Gaussian distorting fields by including higher order $\cal D$ vertices in the diagrams. For example, non-linearities in the lensing potential can be treated by including $\phi^4$, etc.\ interaction vertices.

\section{Feynman Diagrams for Lensing}\label{sec:feynman}

In this section, we focus on how to calculate the effect of gravitational lensing on the observed CMB using Feynman diagrams.
 We will comment on other distortions in \sec{other}.

\subsection{Feynman Rules}\label{subsec:rules}

The starting point is \eqs{RealTay}{deflection}, treating $\phi$ as a Gaussian field with power spectrum $C^{\phi\phi}_\bfl$, and working in the flat-sky approximation. The distortion matrix for gravitational lensing is
\begin{align} \label{eq:lensing}
  D^\text{Lensing}_{(\bfl,\bfm)} &= R_{(-\bfl,\bfm)}\,(2\pi)^2 \delta^2(\bfl-\bfm-\cP) 
  \nn & \quad \times
 \exp\Big[ -\intk\,(\bfk \sdt \bfm)\, \phi_\bfk \Big] 
\,,\end{align}
which generalizes \eq{lensD} to include mixing of the polarization fields.
The exponential contains terms with $n=0,1,2,\ldots$ \ $\phi$ fields,
\begin{align}
  & \exp\Big[ -\intk\,(\bfk \sdt \bfm)\, \phi_\bfk \Big] 
  \nn & \quad
  = \sum_n \frac{1}{n!} \prod_{i=1}^n \Big[-
   \int\! \frac{d^2 \bfk_i}{(2\pi)^2}\,(\bfk_i \sdt \bfm)\, \phi_{\bfk_i} \Big]
\,.
\label{eq:lensingexp}
\end{align}
In \eq{lensing}, $\cP = \sum_{i=1}^n \bfk_i$ gives the total momentum of all the $\phi$ fields for each term in the expansion of the exponential. The mixing of $E$ and $B$ polarizations is described by the rotation matrix
\begin{align}
 R_{(\bfl,\bfm)} 
 &=
 \begin{pmatrix}
   1 & 0 & 0 \\
   0 & \cos 2\varphi(\bfl,\bfm) & \sin 2\varphi(\bfl,\bfm) \\
   0 & -\sin 2\varphi(\bfl,\bfm) & \cos 2\varphi(\bfl,\bfm)
 \end{pmatrix}
\label{eq:rmatrix}
\,,\end{align}
where $\varphi(\bfl,\bfm)=\varphi_\bfl -\varphi_\bfm$ is the angle between $\bfl$ and $\bfm$.  There is no mixing in the lowest order $n=0$ term in \eq{lensing}, which contains no power of the lensing field $\phi$, and reduces to $(2 \pi)^2 \delta^2(\bfl - \bfm)$ since $R_{(-\bfl,\bfl)}=1$ is the identity matrix. 
For calculations, it is convenient to write
\begin{align}
\cos 2\varphi(\bfl,\bfm) &= 2\, \frac{(\bfl \sdt \bfm)^2}{\bfl^2 \bfm^2} - 1
\,,\nn
\sin 2\varphi(\bfl,\bfm) &= 2\, \frac{(\bfl \sdt \bfm)(\bfl * \bfm)}{\bfl^2 \bfm^2}
\,,\end{align}
with $\bfl * \bfm = \epsilon^{ij} \bfl^i \bfm^j$, where our convention for the anti-symmetric tensor $\epsilon$ is $\epsilon^{12}=1$. Note that $\sin 2\varphi(\bfl,\bfm) = - \sin 2\varphi(\bfm,\bfl)$, so that $\sin 2\varphi(\bfl,\bfm)$ depends on the orientation of the angle between $\bfl$ and $\bfm$. However, neither the sine nor cosine depends on the sign of either $\bfl$ or $\bfm$, since switching the sign shifts $\varphi$ by $2\pi$, so
\begin{align} \label{eq:signflip}
 R_{(\bfl,\bfm)} &=  R_{(-\bfl,\bfm)} =  R_{(\bfl,-\bfm)}=R_{(-\bfl,-\bfm)}
\,.\end{align}
Expanding \eq{lensing} to $\ord{\phi^2}$ and using the delta function to perform the integral in \eq{gen} reproduces \eq{lensD}.

\begin{figure*}
\raisebox{-0.5\height}{
\psfrag{labelone}[c][c]{$\bfl$}
\includegraphics[width=2.1cm]{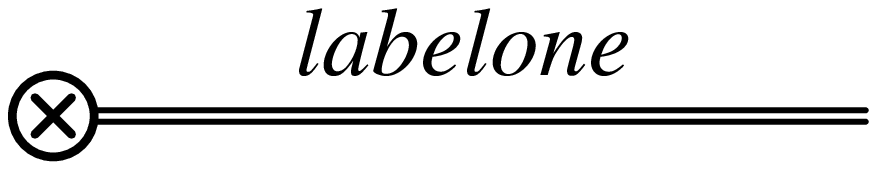}}\quad $=$\quad
\raisebox{-0.5\height}{\psfrag{labelone}[c][c]{$\bfl$}\includegraphics[width=2.1cm]{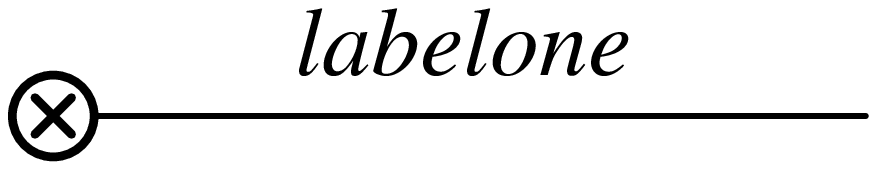}} $\ +$\quad
\raisebox{-0.28\height}{\psfrag{labelone}[c][c]{$\bfk_1$} \psfrag{labeltwo}[c][c]{$\bfm$}
\includegraphics[width=2.3cm]{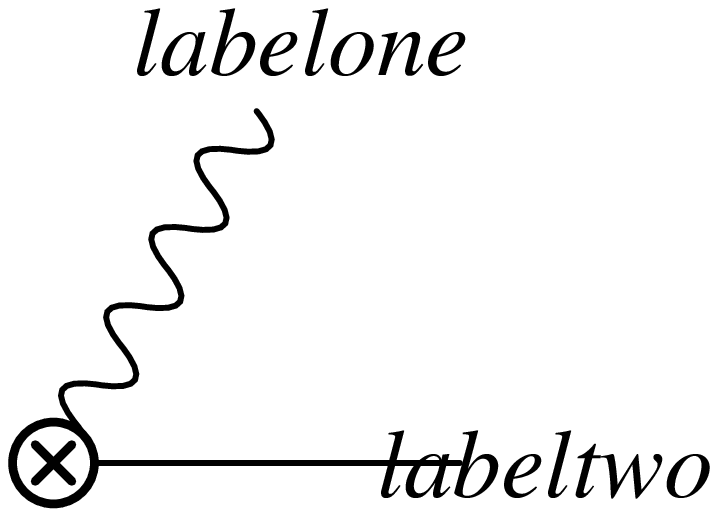}}
\hspace{-0.4cm} $+$\quad
\raisebox{-0.27\height}{\psfrag{labelone}[c][c]{$\bfk_2$}\psfrag{labeltwo}[c][c]{$\bfk_1$} \psfrag{labelthree}[c][c]{$\bfm$}
\includegraphics[width=2.3cm]{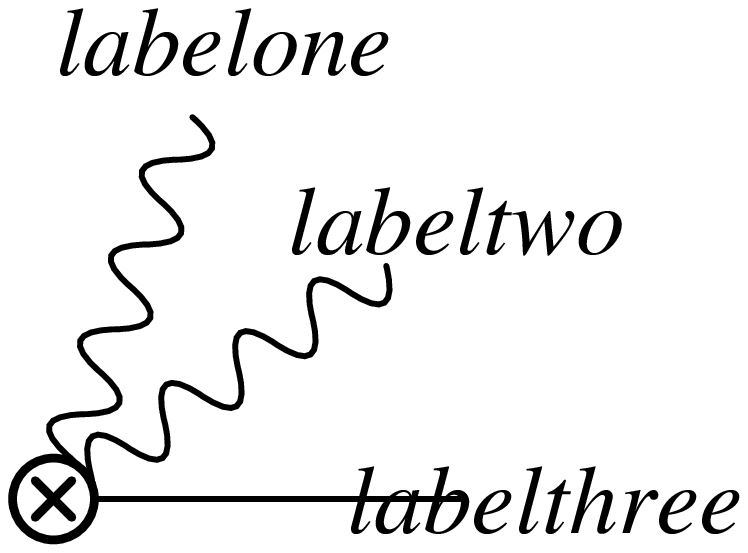}}
\hspace{-0.2cm} $+\  $ \ldots
\caption{Expansion of the lensed field (double line) in terms of the unlensed field (solid line) and powers of the lensing potential $\phi$ (wiggly line).
\label{fig:expand}}
\end{figure*}

\begin{figure}[t]
\begin{tabular}{cl}
\raisebox{-0.5\height}{
\psfrag{labelone}[r][r]{$x$}
\psfrag{labeltwo}[l][l]{$y$}
\psfrag{labelthree}[c][c]{$\bfm$}
\includegraphics[width=4.5cm]{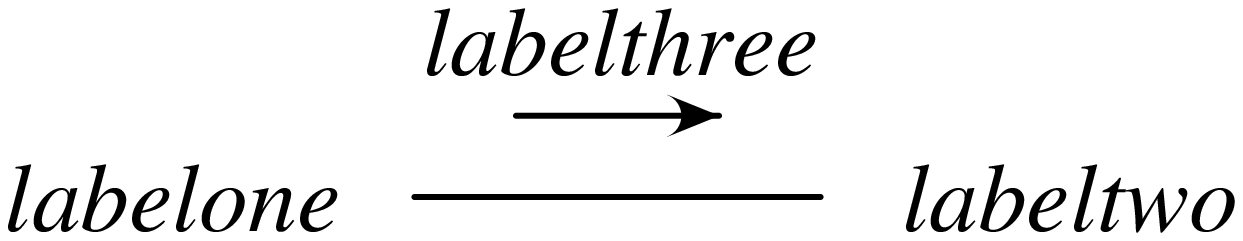}} & $C^{xy}_\bfm$ \\
\raisebox{-0.5\height}{
\psfrag{labelone}[c][c]{$\bfk$}\includegraphics[width=1.8cm]{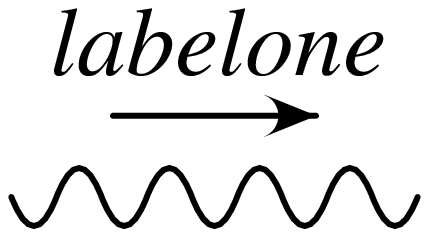}} & $C^{\phi\phi}_\bfk$ \\[1cm]
\raisebox{-0.5\height}{
\psfrag{labelone}[r][r]{$\bfl,x$}
\psfrag{labeltwo}[c][c]{$\bfm,y$}
\psfrag{labelthree}[c][c]{$\bfk_1$}
\psfrag{labelfour}[c][c]{$\bfk_2$}
\psfrag{labelfive}[c][c]{$\bfk_n$}
\includegraphics[width=5cm]{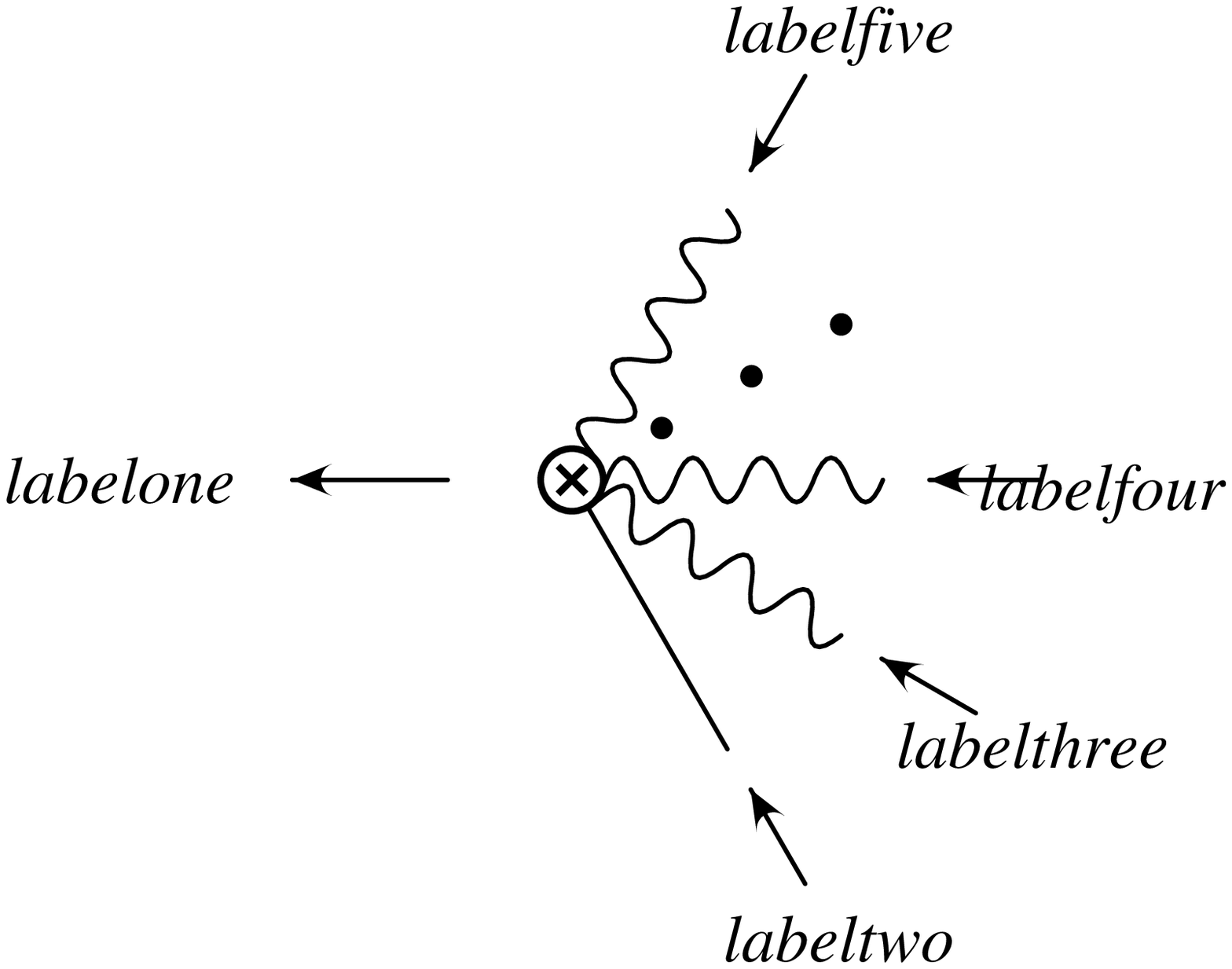}} & $R_{(-\bfl,\bfm)}^{xy} \prod_i (-\bfk_i \sdt \bfm)$ \\
\end{tabular}
\caption{Feynman rules for calculating lensed CMB fields. The labels  $x, y \in \{T,E,B\}$ denote the type of field. For the last diagram, $x \in \{\widetilde T, \widetilde E, \widetilde B\}$ is lensed and $y$ is unlensed. The outgoing momentum $\bfl$ is equal to the total incoming momentum, $\bfl = \bfm + \sum_i \bfk_i$.}
\label{fig:rules}
\end{figure}

Although \eq{lensing} may seem complicated, it can be described by a simple Feynman rule. For an introduction to Feynman diagrams in quantum field theory, see 
e.g.~Ref.~\cite{Brown:1992db,Peskin:1995ev}.
When calculating the average of several CMB modes $\langle \widetilde x_\bfl \widetilde y_\bfk \dots \rangle$ over CMB or LSS realizations, each lensed field $\widetilde x_\bfl$ is expanded as the sum of vertices with one solid line (the unlensed field) and $n=0,1,2,\cdots$ wiggly lines (the lensing field $\phi$) connected to it, as shown in \fig{expand}.  Each vertex in the expansion has momentum $\bfl$ flowing in, and the double line represents the lensed field. The unlensed field and the $\phi$ fields each have momenta flowing into the vertex, given by their label in \fig{expand}. 
The vertex with $n$ wiggly lines arises from the $n^{\rm th}$-order term in the expansion of \eq{lensingexp}, as shown in the third graph in \fig{rules}.  The factor of 
$(-\bfk_i \cdot \bfm)$ for each $\phi$ line is the same factor as in \eq{lensingexp}. The $1/n!$ in \eq{lensingexp} is cancelled by the standard $n!$ combinatorial factor in a vertex with $n$ identical fields from the $n!$ possible permutations of the $\phi$ fields. It is therefore absent from the Feynman rule in Fig.~\ref{fig:rules}. The superscript $xy$ on $R_{(-\bfl,\bfm)}^{xy}$ denotes the $xy$ element of the matrix $R_{(-\bfl,\bfm)}$, which is given by \eq{rmatrix}.

We now discuss how to determine $\langle \widetilde X_1 \ldots \widetilde X_n\rangle$.
Every lensed field $\widetilde X_i$  has an expansion as described in the above paragraph, and shown diagrammatically in Fig.~\ref{fig:expand}. Taking the CMB average involves computing the correlation $\vev{ X_1 \ldots  X_n}$ of the unlensed fields (solid lines) arising from the expansion of the lensed fields using \eq{lensing}. Since the primordial CMB fluctuations are Gaussian, the $n$-point correlator is given by Wick's theorem, which corresponds to joining pairs of primordial fields together to form internal solid lines in all possible ways. The solid line is  the first graph in Fig.~\ref{fig:rules}. It corresponds to the matrix propagator \eq{cmbavg} of the CMB spectrum, and the indices $xy$ are the CMB components $T,E,B$ which are contracted together.

The average over LSS realizations is carried out using \eq{lssavg}. Again, this is accomplished diagrammatically by joining pairs of lensing $\phi$ fields (wiggly lines in Fig.~\ref{fig:rules}) in all possible ways. The lensing spectrum $C^{\phi \phi}_\bfk$ is used for each wiggly line propagator with momentum $\bfk$ flowing through it.

All lines in a diagram have a momentum associated with them, and the total momentum is conserved at each vertex. This is where using the flat-sky approxmation simplifies the formulae, because in this approximation, angular momentum behaves like linear momentum with the conservation law in \eq{flat}, rather with the original angular momentum conservation law in \eq{full}. The momentum $\bfk$ of an internal line is unconstrained and integrated over with $d^2\bfk/(2\pi)^2$. 
Many of these  integrals are done with the help of the momentum-conserving delta-functions at each vertex, so that the only remaining integrals are over closed loops, as is well-known from the usual treatment of Feynman graphs.

Non-gaussianities in the primordial CMB fluctuations, or in LSS can be included as additional higher order vertices, i.e.~involving more than two  $X$ and $\phi$ fields.  These additional vertices result in additional graphs, but the Feynman diagram method can still be used. We restrict our attention to the Gaussian case for the rest of this paper.

In summary, the $n$-point lensed correlation function $\langle \widetilde X_1 \ldots \widetilde X_n \rangle$ is given by the sum of all diagrams using the Feynman rules in 
Fig.~\ref{fig:rules}. There are no external (unpaired) solid lines if the average is taken over CMB realizations, and no external $\phi$ lines if the average is taken
over LSS realizations. One important point to note is that the Feynman graphs will often have symmetry factors, which are the same as the usual ones in quantum field theory.

\subsection{Lensing Filters}\label{subsec:filters}

In the presence of lensing, the average of CMB modes $x$ and $y$ over CMB realizations takes the following form
\begin{align}\label{eq:2point}
 \langle \widetilde x_{\bfl} \widetilde y_{\bfL-\bfl}\rangle_\text{CMB} 
 &= 
  (2\pi)^2 \delta^2(\bfL)  \big[C^{xy}_\bfL + \dots\big]
 \\ & \quad 
 + \big[\ff{xy}{\phi,0}{(\bfl,\bfL-\bfl)} + \ff{xy}{\phi,1}{(\bfl,\bfL-\bfl)} + \dots \big] \phi_\bfL 
 \nn & \quad
+ \intm\, \ff{xy}{\phi\phi,0}{(\bfl,\bfL-\bfl,\bfm)} \phi_{\bfL-\bfm} \phi_\bfm 
 + \dots
\,.\nonumber\end{align}
Treating $x$ and $y$ as components of the vector $X=\begin{pmatrix} T & E & B \end{pmatrix}$, \eq{2point} can be viewed as a matrix equation and the coefficients $f$, known as filters, as $3\times 3$ matrices. The first index $i$ on $f^{(i,j)}$ denotes the number of external $\phi$ fields in the average, so that $i=\phi$ is one external $\phi$ field, $i=\phi\phi$ is two external $\phi$ lines, etc. The second index $j$ is the order in the lensing power spectrum $C^{\phi\phi}$ of the term. In Feynman diagram language, this is the expansion of the $xy$ vertex in powers of an external background $\phi$ field.

\begin{figure}
\raisebox{-0.5\height}{
\psfrag{labelone}[c][c]{\raisebox{2mm}{$x,\bfl$}}
\psfrag{labeltwo}[c][c]{\raisebox{2mm}{$y,\bfl^\prime$}}
\psfrag{labelthree}{$\bfl+\bfl^\prime$}
\includegraphics[width=3cm]{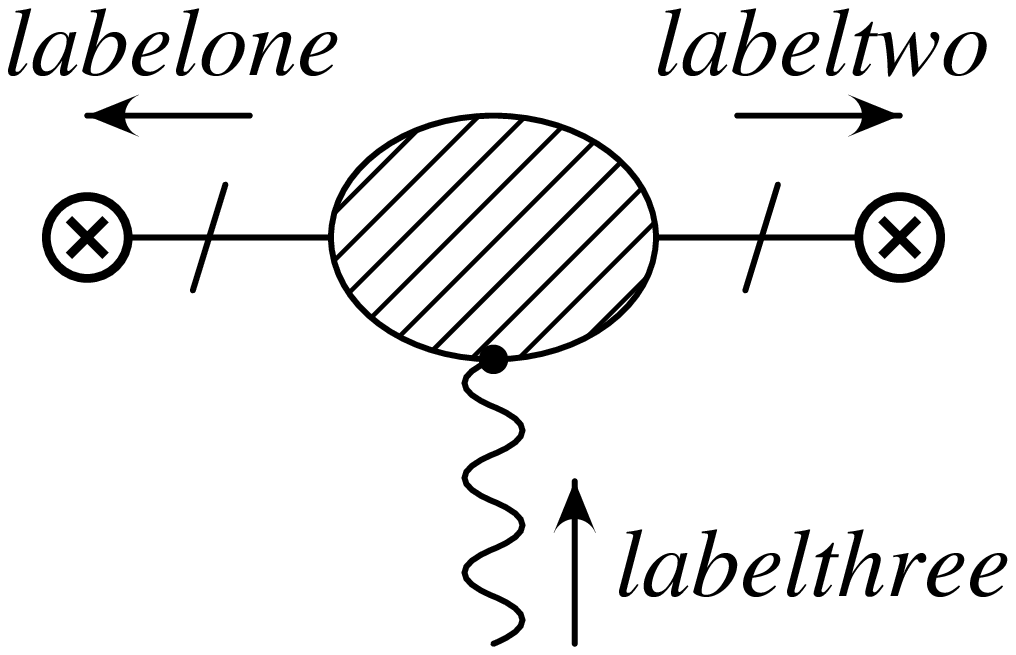}} $=$
\raisebox{-0.5\height}{\includegraphics[width=0.11\textwidth]{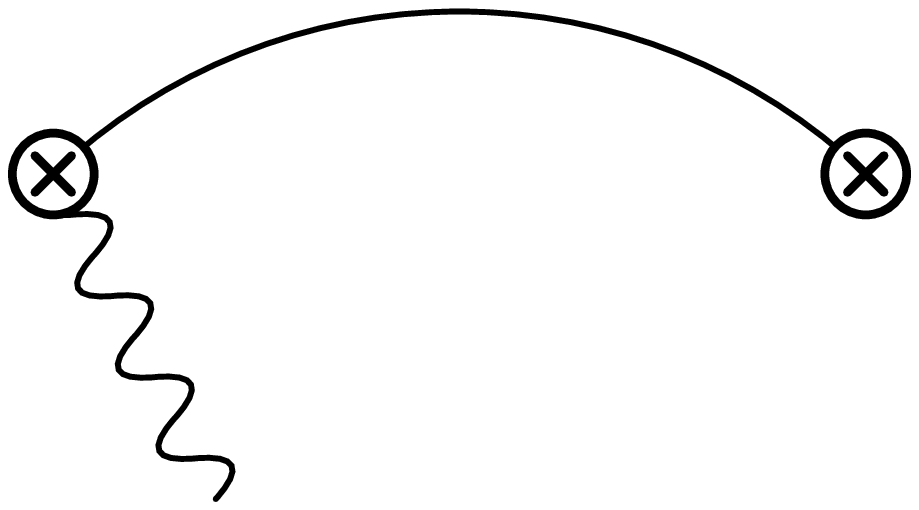}} $+$
\raisebox{-0.5\height}{\includegraphics[width=0.11\textwidth]{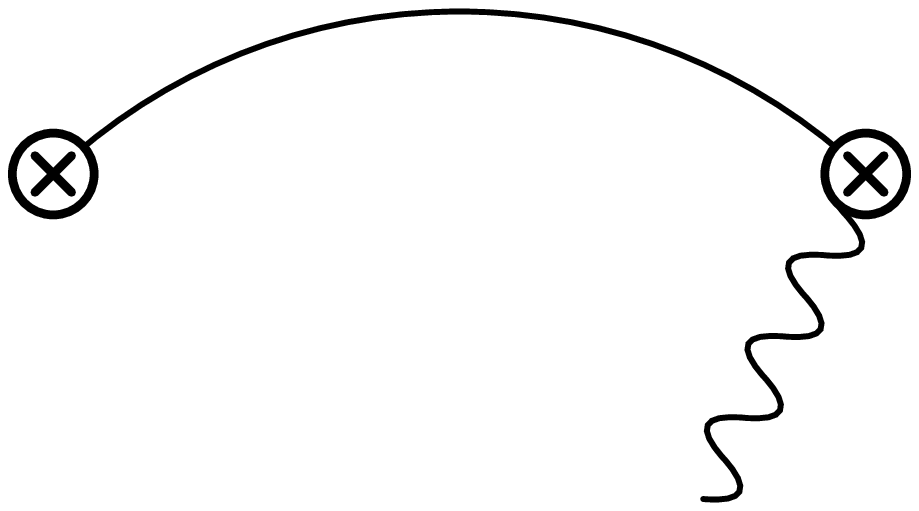}}
\caption{Diagrams contributing to $f^{(\phi,0)xy}_{(\bfl,\bfl')}$.
\label{fig:f0}}
\end{figure}

We now calculate the filters $\ff{}{\phi,0}{}$, $\ff{}{\phi,1}{}$ and $\ff{}{\phi\phi,0}{}$, which are the ingredients necessary to determine the noise up to $\ord{\phi^4}$. 
The filter $\ff{}{\phi,0}{}$ receives contributions from the diagrams shown on the r.h.s.~in \fig{f0}, where the lensing field $\phi$ attaches to either vertex. The fields $x$ and $y$ have outgoing momentum $\bfl$ and $\bfl^\prime$, so by momentum conservation the incoming momentum along the wiggly line is $\bfl+\bfl^\prime$. Using the rules in \fig{rules}, the sum of these diagrams is
\begin{align} 
\ff{}{\phi,0}{(\bfl,\bfl')}&=R_{(-\bfl,-\bfl')} C_{\bfl'}  (\bfl+\bfl')\cdot \bfl' 
\nn & \quad
+  C_\bfl R^T_{(-\bfl',-\bfl)} (\bfl+\bfl')\cdot \bfl 
\label{eq:ff0}
\,.\end{align}
The graphical representation of the $xy\phi$ vertex $\ff{}{\phi,0}{}$  is the shaded blob on the l.h.s.~of \fig{f0}. The slashes on the external solid lines representing the unlensed fields $x$ and $y$ is a reminder that the rule for the graph \emph{does not} include propagators (spectra) for the external lines. There is only one propagator in  $\ff{}{\phi,0}{}$, as can seen explicitly from the r.h.s. of \fig{f0} and in \eq{ff0}.

\begin{figure}
\raisebox{-0.5\height}{\includegraphics[width=0.11\textwidth]{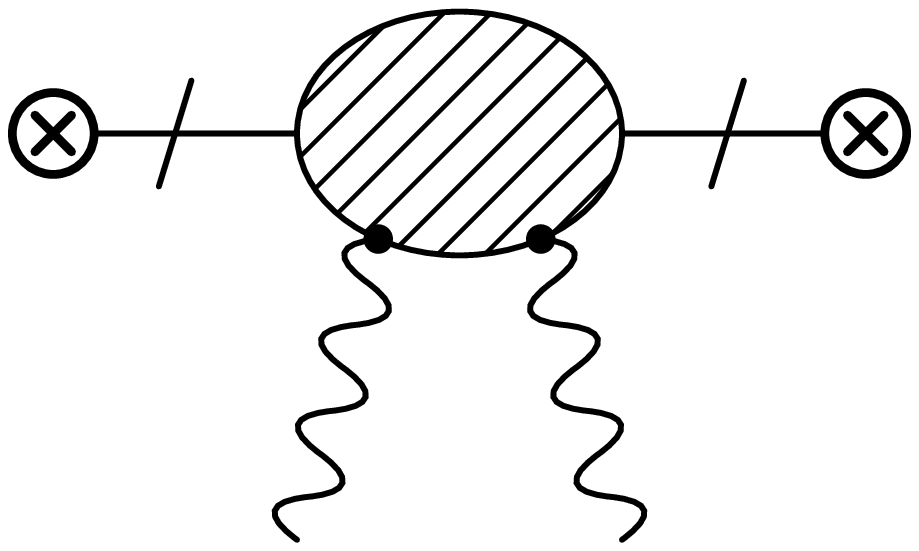}} $=$
\raisebox{-0.5\height}{\includegraphics[width=0.1\textwidth]{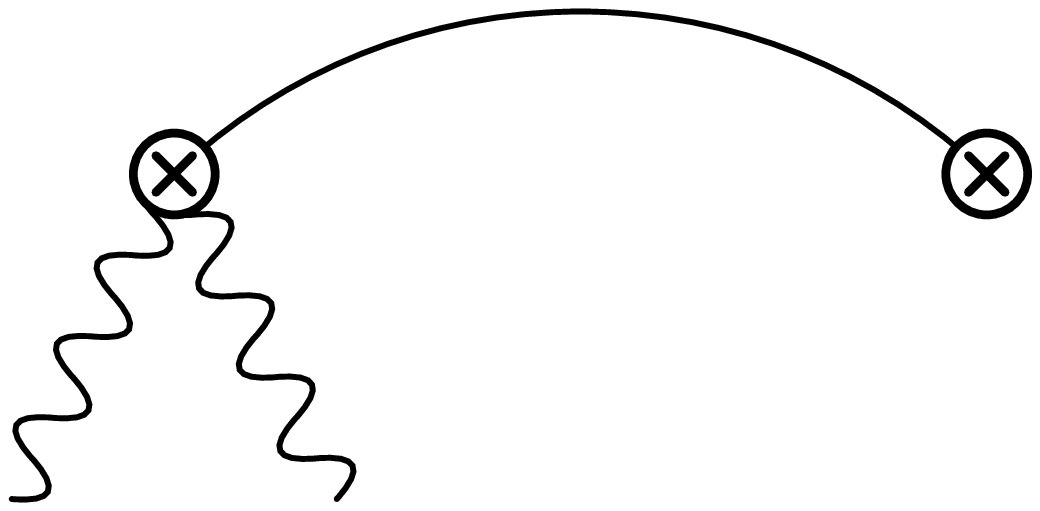}} $+$\!
\raisebox{-0.5\height}{\includegraphics[width=0.1\textwidth]{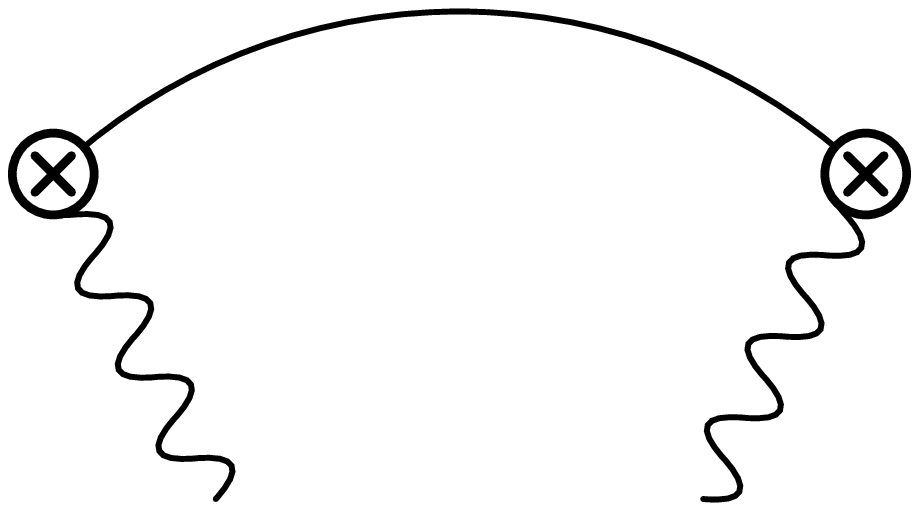}} $+$
\raisebox{-0.5\height}{\includegraphics[width=0.1\textwidth]{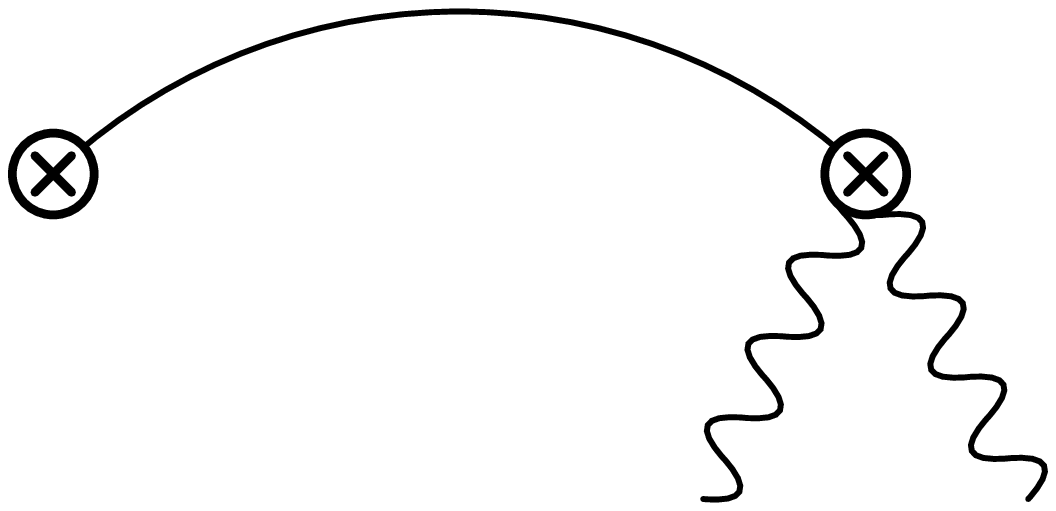}}
\caption{Diagrams contributing to $f^{(\phi\phi,0)}$.
\label{fig:g}}
\end{figure}

\begin{figure}
\begin{tabular}{ccccccc}
\raisebox{-0.7\height}{
\includegraphics[width=0.1\textwidth]{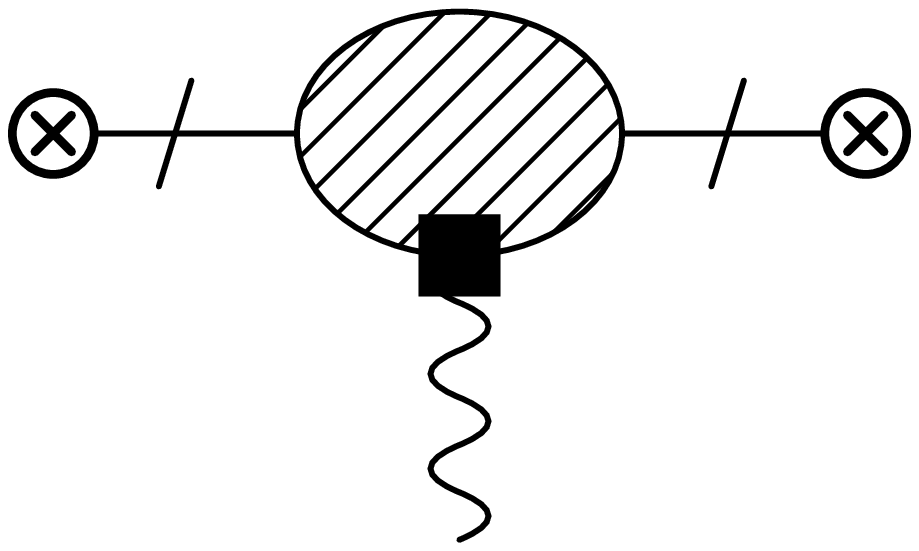}} & $=$ &
\raisebox{-0.5\height}{\includegraphics[width=0.098\textwidth]{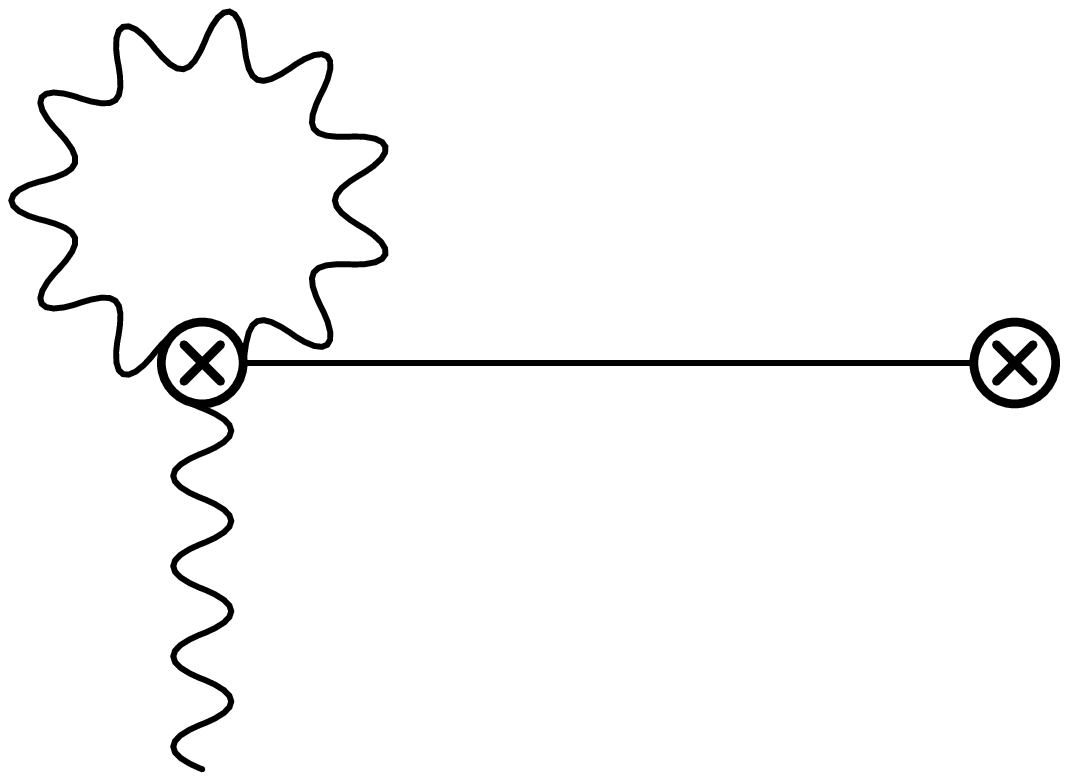}} & $+$ &
\raisebox{-0.5\height}{\includegraphics[width=0.098\textwidth]{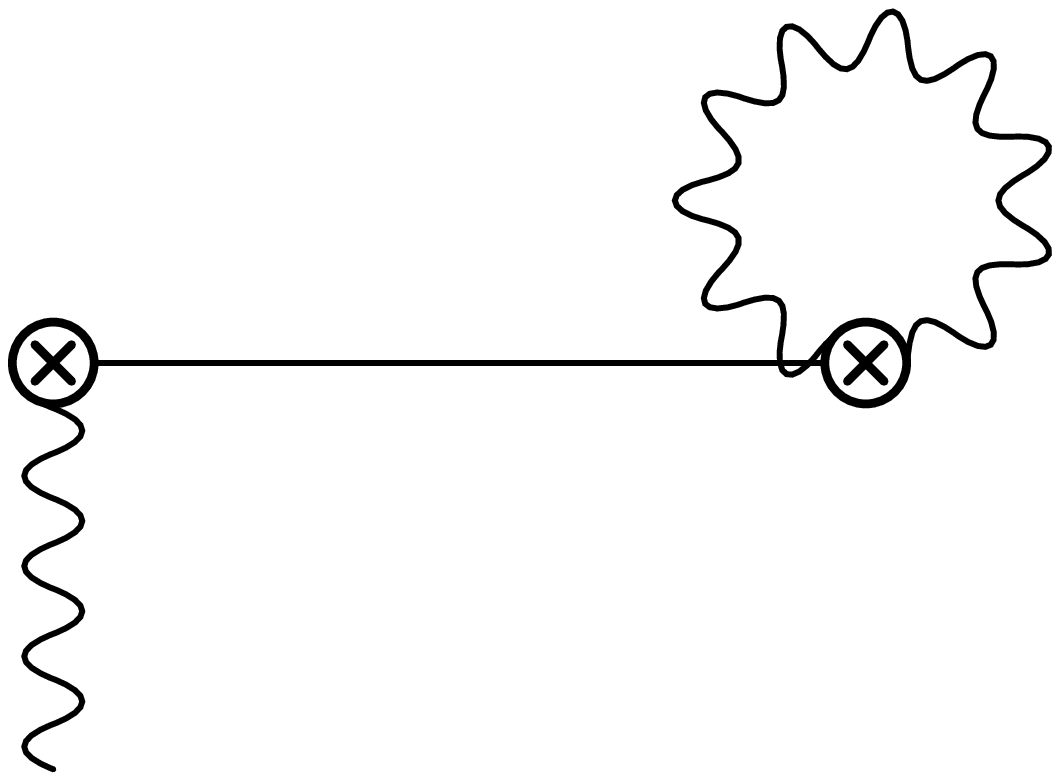}} & $+$ &
\raisebox{-0.55\height}{\includegraphics[width=0.085\textwidth]{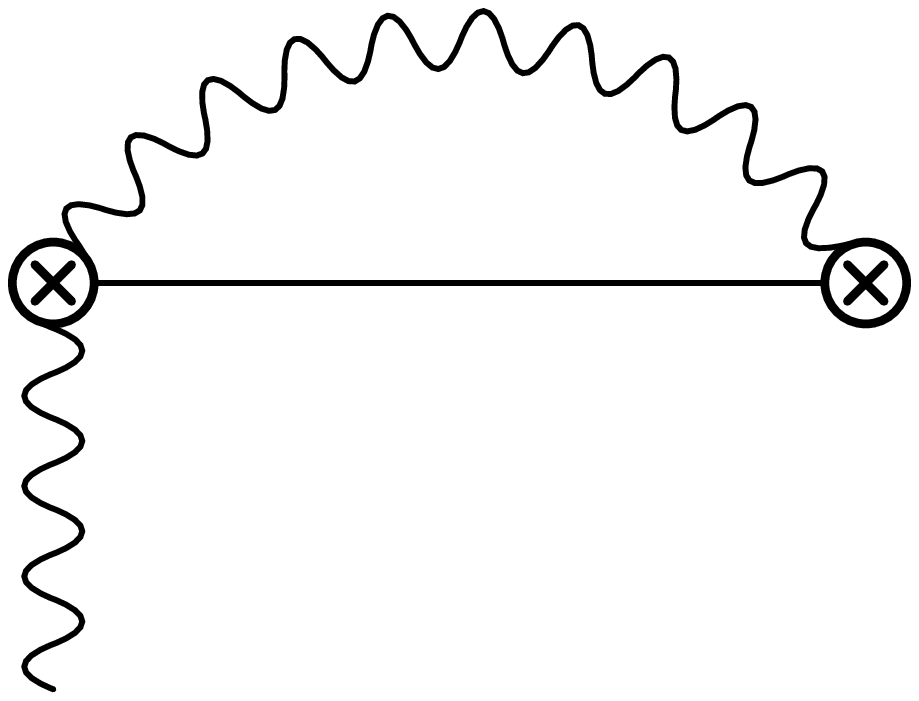}} \\
&& (a) && (b) && (c)
\end{tabular}
\caption{Diagrams contributing to $f^{(\phi,1)}$. Mirror graphs where the external $\phi$ line comes out of the second vertex have not been shown.
\label{fig:f1}}
\end{figure}

The diagrams contributing to $\ff{}{\phi\phi,0}{}$, shown in \fig{g}, involve two external wiggly lines corresponding to the two $\phi$ fields. One of the momenta of the $\phi$ fields is arbitrary, and we will take it to be $\bfm$. The other is fixed to be $\bfl + \bfl^\prime - \bfm$ by momentum conservation. The first and third diagram on the r.h.s.~in \fig{g} have a symmetry factor of $1/2$. We symmetrize the middle graph under $\bfm \leftrightarrow \bfl + \bfl^\prime - \bfm$, such that in the graphical representation of the $f^{(\phi\phi,0)}$ vertex on the l.h.s., the two wiggly lines are identical. Adding up these contributions gives
\begin{widetext}
\begin{align} \label{eq:fphiphi0}
f^{(\phi\phi,0)}_{(\bfl,\bfl',\bfm)}&=\frac{1}{2} \Big\{R_{(-\bfl,-\bfl')} C_{\bfl'} (\bfm\sdt \bfl')[(\bfl+\bfl'-\bfm)\sdt \bfl'] - R_{(-\bfl,\bfl-\bfm)}  C_{\bfl-\bfm} R^T_{(-\bfl',\bfm-\bfl)}  [\bfm\sdt (\bfl-\bfm)][(\bfl+\bfl'-\bfm)\sdt (\bfl-\bfm)] 
\nonumber \\  
&\quad + C_{\bfl} R^T_{(-\bfl',-\bfl)}  (\bfm\sdt \bfl)[(\bfl+\bfl'-\bfm)\sdt \bfl] - R_{(-\bfl,\bfm-\bfl')}  C_{\bfm-\bfl'} R^T_{(-\bfl',\bfl'-\bfm)}  [\bfm\sdt (\bfl'-\bfm)][(\bfl+\bfl'-\bfm)\sdt (\bfl'-\bfm)]  \Big\}
.\end{align}

We now consider $f^{(\phi,1)}$, which is the $\ord{\phi^2}$ correction to $f^{(\phi,0)}$. The diagrams, shown in \fig{f1},  involve an additional internal wiggly line whose loop momentum $\bfm$  is unconstrained and gets integrated over, yielding
\begin{align}
\!\!f^{(\phi, 1)}_{(\bfl,\bfl^\prime)} &=  \!\!\intm\Big\{ R_{(-\bfl,\bfm-\bfl^\prime)} C_{\bfm-\bfl^\prime} R^T_{(-\bfl^\prime,\bfl^\prime-\bfm)} \left[\bfm \sdt \left(\bfl^\prime\!-\!\bfm\right)\right]^2
\left[\left(\bfl^\prime\!-\!m\right) \sdt \left(\bfl\!+\!\bfl^\prime\right)\right] C^{\phi \phi}_\bfm 
\!-\! R_{(-\bfl,-\bfl^\prime)} C_{\bfl^\prime} \left( \bfm \sdt \bfl^\prime \right)^2 
\left[ \bfl^\prime \sdt \left(\bfl\!+\!\bfl^\prime\right)\right] C^{\phi \phi}_\bfm
\nn & \quad
+R_{(-\bfl,\bfl-\bfm)} C_{\bfl-\bfm} R^T_{(-\bfl^\prime,\bfm-\bfl)} \left[\bfm \sdt \left(\bfl\!-\!\bfm\right)\right]^2
\left[\left(\bfl\!-\!\bfm\right)\sdt \left(\bfl\!+\!\bfl^\prime\right)\right] C^{\phi \phi}_\bfm
\!-\! C_{\bfl} R^T_{(-\bfl^\prime,-\bfl)} \left( \bfm \sdt \bfl \right)^2 
\left[ \bfl \sdt \left(\bfl\!+\!\bfl^\prime\right)\right] C^{\phi \phi}_\bfm \Big\}
\,.\end{align}
\end{widetext}
 $f^{(\phi,1)}$ is denoted by a shaded blob with a small black square, as depicted in \fig{f1}.  The black square distinguishes the $f^{(\phi,1)}$ vertex from the $f^{(\phi,0)}$ vertex. It is $\ord{\phi^2}$ suppressed and contains an internal integral over loop momentum.

\subsection{Organizing the Expansion using Lensed Spectra}\label{subsec:expansion}

\begin{figure*}
\begin{tabular}{ccccccccc}
\raisebox{-0.2\height}{\includegraphics[width=0.12\textwidth]{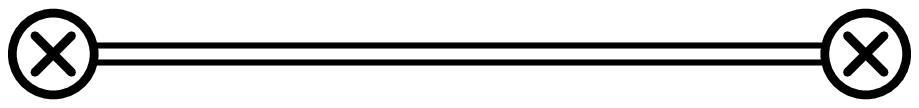}} & $=$ &
\raisebox{-0.2\height}{\includegraphics[width=0.12\textwidth]{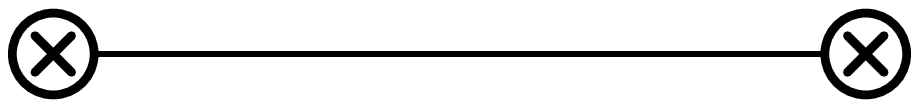}} & $+$ &
\raisebox{-0.05\height}{\includegraphics[width=0.14\textwidth]{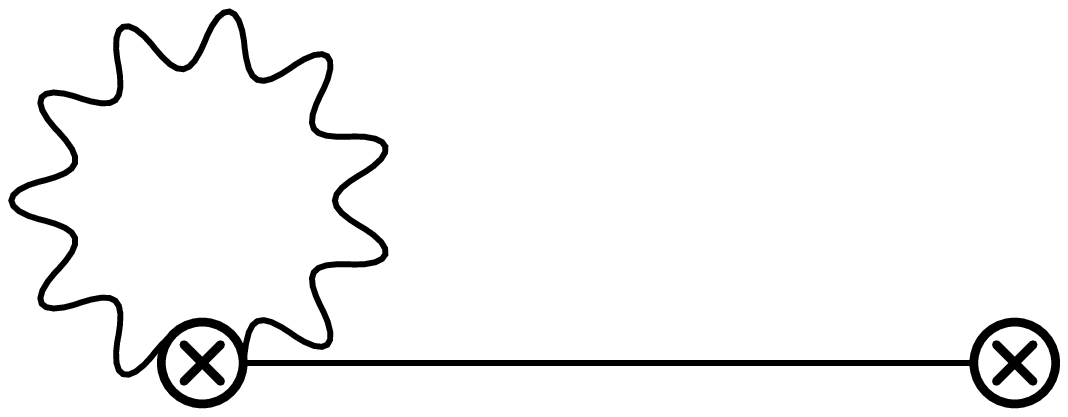}} & $+$ &
\raisebox{-0.05\height}{\includegraphics[width=0.14\textwidth]{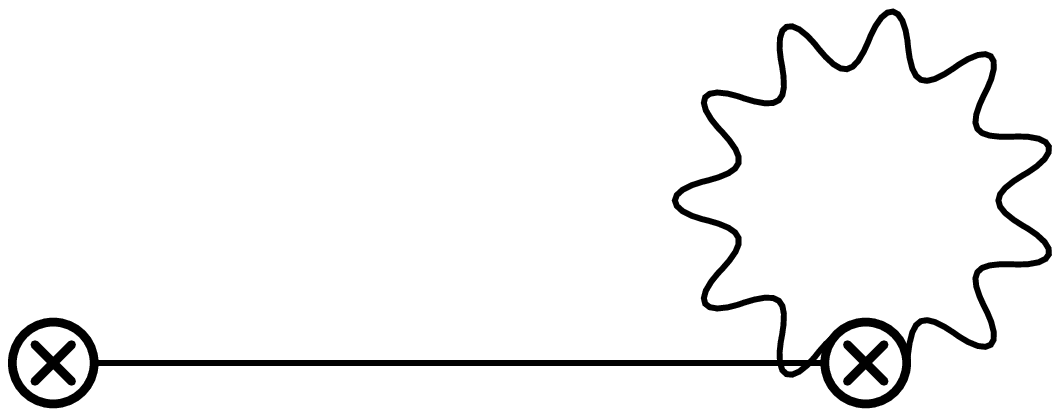}} & $+$ &
\raisebox{-0.05\height}{\includegraphics[width=0.12\textwidth]{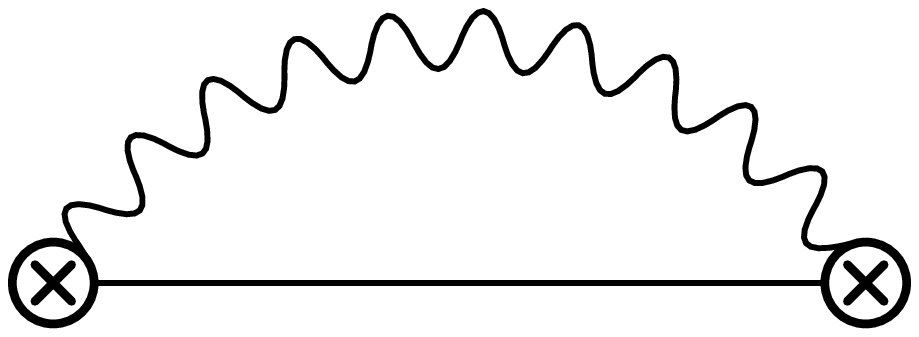}} \\[5pt]
&& && (a) && (b) && (c)
\end{tabular}
\caption{The two point function $\widetilde C_\bfl$ using the lensed correlator is depicted by a double line. Its expression in terms of the unlensed correlator $C_\bfl$ is given to order $\ord{\phi^2}$  by the sum of the unlensed propagator $C_\bfl$ and diagrams (a), (b), (c).
\label{fig:7}}
\end{figure*}

\begin{figure}
\begin{tabular}{ccc}
\raisebox{-0.5\height}{\includegraphics[width=0.14\textwidth]{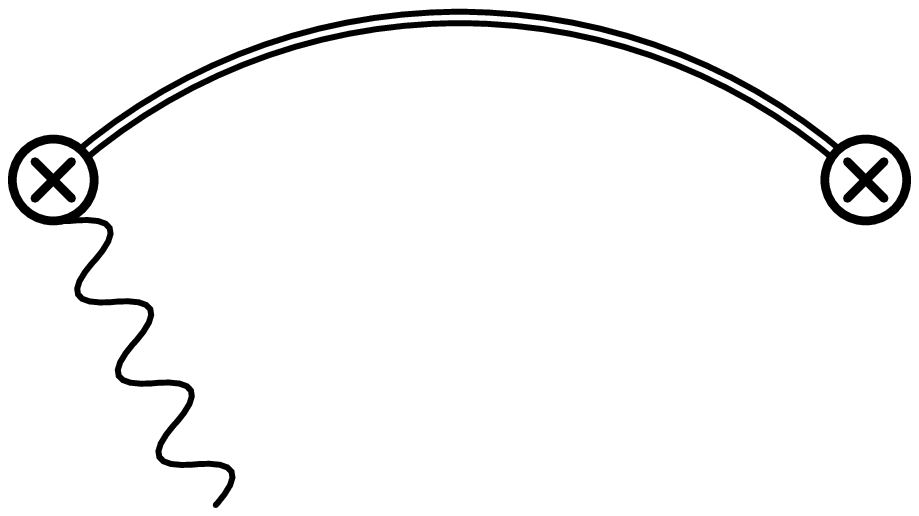}} & $+$ &
\raisebox{-0.5\height}{\includegraphics[width=0.14\textwidth]{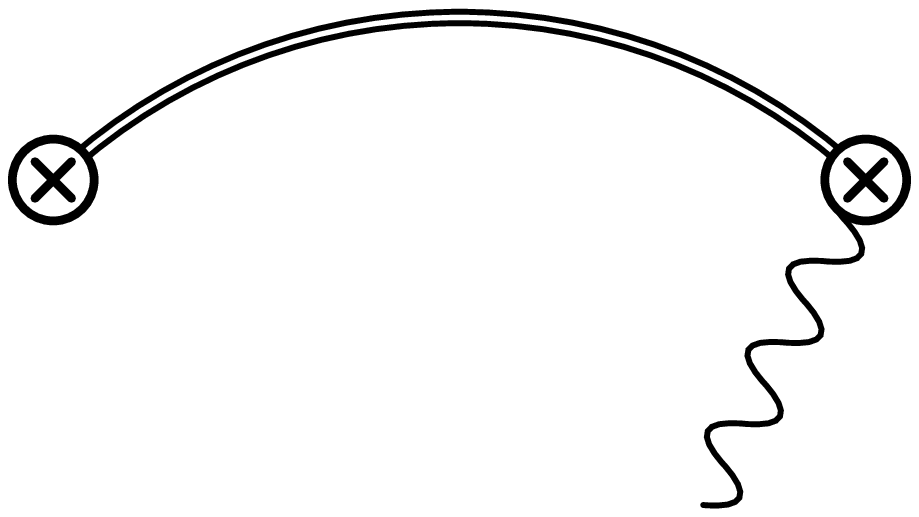}}
\end{tabular}
\caption{Graphs contributing to the lensed filter $\widetilde f^{(\phi,0)}$.
\label{fig:8}}
\end{figure}

In \eq{2point}, we can replace the unlensed power spectra $C^{xy}$ by lensed ones $\widetilde C^{xy}$ on the first line and in the filters $f^{(\phi,0)}$ and $f^{(\phi\phi,0)}$. We will denote the corresponding filters by 
$\widetilde f^{(\phi,0)}$ and $\widetilde f^{(\phi\phi,0)}$, such that 
\begin{align}\label{eq:2point_lensed}
 \langle \widetilde x_{\bfl} \widetilde y_{\bfL-\bfl}\rangle_\text{CMB} 
 &= 
  (2\pi)^2 \delta^2(\bfL)  \widetilde C^{xy}_\bfL
 \\ & \quad 
 + \big[\widetilde f^{(\phi,0)xy}_{(\bfl,\bfL-\bfl)} + \widetilde f^{(\phi,1)xy}_{(\bfl,\bfL-\bfl)} + \dots \big] \phi_\bfL 
 \nn & \quad
+ \intm\, \widetilde f^{(\phi\phi,0)xy}_{(\bfl,\bfL-\bfl,\bfm)} \phi_{\bfL-\bfm} \phi_\bfm 
 + \dots
\,,\nonumber\end{align}
where $\widetilde f^{(\phi,1)}$ is defined so that \eq{2point_lensed} is in agreement with \eq{2point}.
This change amounts to absorbing some terms that are formally higher order in $\phi$ into the lensed filters, and thus reorganizes the lensing expansion. In particular, the lensed spectrum captures all of the higher-order contributions to the first line.
The lensed propagator $\widetilde C^{xy}$ is depicted by a double line, and is given to order $\ord{\phi^2}$ by the sum of diagrams in \fig{7}. The lensed filter $\widetilde f^{(\phi,0)}$ is given by \eq{ff0} with $C_\bfl \to \widetilde C_\bfl$, i.e.~by the graphs shown in \fig{8}. Similarly, the lensed filter $\widetilde f^{(\phi\phi,0)}$ is given by \eq{fphiphi0} with $C_\bfl \to \widetilde C_\bfl$, since it is the lowest order contribution with two external $\phi$'s.  The definition of the lensed filter $\widetilde f^{(\phi,1)}$ is more involved.

Using \fig{7} for the lensed correlator in \fig{8}, one sees that parts of the unlensed filter $f^{(\phi,1)}$ are now already included in the lower order lensed filter $\widetilde f^{(\phi,0)}$. The graphs in \fig{f1}(a) and (b) are completely included. \fig{f1}(c) is not the same as the graph obtained by using \fig{7}(c) for the lensing propagator in the graphs of \fig{8}, because the momentum labels on the internal lines are different in the two graphs. The difference between these two expressions is the lensed filter 
$\widetilde f^{(\phi,1)}$,
\begin{align} \label{eq:f1}
\widetilde f^{(\phi,1)}_{(\bfl, \bfl')} 
&=  \!\!\intm \big\{R_{(-\bfl,\bfm-\bfl')} \widetilde C_{\bfm-\bfl'} R^T_{(-\bfl',\bfl'-\bfm)} [\bfm\sdt (\bfl'\!-\!\bfm)]^2
\nn & \quad
+ R_{(-\bfl,\bfl-\bfm)} \widetilde C_{\bfl-\bfm} R^T_{(-\bfl', \bfm-\bfl)} [\bfm\sdt (\bfl\!-\!\bfm)]^2 \big\}
\nn & \quad \times
C^{\phi \phi}_\bfm [-\bfm\sdt (\bfl+\bfl')]
\,.\end{align}
In deriving \eq{f1}, one finds two contributions: a piece from changes in the propagator and a piece from changes in the angles in the rotation $R_{(\bfl,\bfm)}$. The latter yields
\begin{align}
 &\big[ R_{(-\bfl,\bfm-\bfl^\prime)}  -R_{(-\bfl,-\bfl^\prime)} R_{(\bfl^\prime ,\bfm-\bfl^\prime)} \big]
C_{\bfm-\bfl^\prime} 
\nn & \quad \times
R^T_{(-\bfl^\prime,\bfl^\prime-\bfm)}
\left[\bfm \sdt \left(\bfl^\prime-\bfm\right)\right]^2
\left[\bfl^\prime \sdt \left(\bfl+\bfl^\prime\right)\right] C^{\phi \phi}_\bfm 
\nn & \quad
+ R_{(-\bfl,\bfl-\bfm)} C_{\bfl-\bfm} \big[ R^T_{(-\bfl^\prime,\bfm-\bfl)} \!-\! R^T_{(\bfl,\bfm-\bfl)}
\nn & \quad \times
R^T_{(-\bfl^\prime,-\bfl)} \big] \left[\bfm \sdt \left(\bfl-\bfm\right)\right]^2
\left[\bfl \sdt \left(\bfl+\bfl^\prime\right)\right] C^{\phi \phi}_\bfm
\,,\end{align}
and vanishes due to the identity 
\begin{align}
R_{({\bf c},-{\bf a})} R_{({\bf a},{\bf b})} &= R(2 \varphi_{\bf c}-2 \varphi_{\bf a}) R(2\varphi_{\bf a}-2 \varphi_{\bf b}) 
\nn &
=R(2\varphi_{\bf c}-2\varphi_{\bf b})=R_{({\bf c},{\bf b})}
\,,\end{align}
which follows from \eq{signflip} and the additive property of rotations.

\section{Other Distorting Fields}\label{sec:other}

The formalism presented in  \sec{feynman} for lensing can also be used to study other cosmological effects such as patchy reionization~\cite{2009PhRvD..79d3003D,2000ApJ...534..533C}, and cosmological rotation of the plane of polarization of the CMB either due to primordial magnetic fields~\cite{Kosowsky:1996yc,2005PhRvD..71d3006K, 2012PhRvD..86l3009Y} or due to a parity-violating Chern-Simons type coupling~\cite{2009PhRvL.102k1302K, 2009PhRvD..79l3009Y, 2009PhRvD..80b3510G}. In general, the quadratic estimator can be used to study any distortion which depends on the line of sight~\cite{Yadav:2009za}, and most of them provide a handle on instrumental systematics~\cite{2003PhRvD..67d3004H}. Below, we discuss cosmological rotation and patchy reionization, and derive the corresponding Feynman rules. Current studies of rotation and patchy reionization using the quadratic estimator are restricted to the leading order estimator noise $N^{(0)}_L$. Using the formalism of this paper, it is straightforward to investigate the higher order noise.

\subsection{Cosmological Rotation}

$CP$ is violated by weak interactions, and it must be violated in the early universe in order to give rise to the baryon asymmetry. This provides a motivation to investigate the existence
of $CP$-violating interactions involving cosmologically
evolving pseudoscalar fields. For example, a Chern-Simons coupling of the form $a F_{\mu \nu}\widetilde F^{\mu \nu}$~\cite{1998PhRvL..81.3067C,2009PhRvL.103e1302P} violates $CP$. It
has been shown that such a term can rotate the polarization
vector of linearly polarized light by an angle of rotation
$d\alpha =2  d \tau\, \dot a$ 
during a conformal time 
$d\tau$. The fluctuations in the scalar field $a$ will then be imprinted in the rotation angle $\alpha$ of the polarization. The shift symmetry of the Lagrangian implies that the field $a$ is classically massless and that the quantum fluctuations frozen in the field during inflation will result in Gaussian perturbations for the rotation $\alpha$ with a nearly scale invariant spectrum.

\begin{figure}[b]
\begin{tabular}{cl}
\raisebox{-0.5\height}{
\psfrag{labelone}[r][r]{$\bfl,x$}
\psfrag{labeltwo}[c][c]{$\bfm,y$}
\psfrag{labelthree}[c][c]{$\bfk_1$}
\psfrag{labelfour}[c][c]{$\bfk_2$}
\psfrag{labelfive}[c][c]{$\bfk_n$}
\includegraphics[width=5cm]{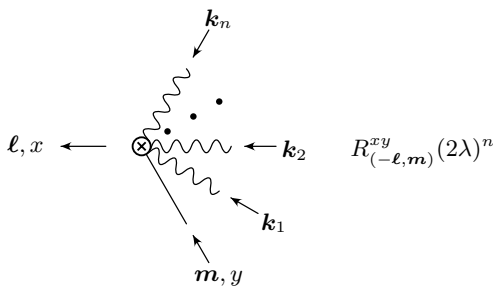}} & $R_{(-\bfl,\bfm)}^{xy} (2\lambda)^n$ \\
\end{tabular}
\caption{Feynman rule for calculating the effect of rotation on the CMB. The primordial CMB component is labelled $y$ and the rotated field $x$. The Feynman rule for the vertex only depends on the number $n$ of rotation fields $\alpha$ (wiggly lines) and not on their momenta $\bfk_i$.}
\label{fig:rotation}
\end{figure}

We will continue using the notation introduced for lensing, denoting the observed (rotated) Stokes parameters with a tilde $\widetilde Q$, $\widetilde U$.
The rotated Stokes parameters are related to the primordial Stokes parameters by
\begin{equation}
[\widetilde Q({\bf n}) \pm i \widetilde U({\bf n})]=e^{\pm 2i \alpha(\bf n)}[Q({\bf n})\pm i U({\bf n})] 
\,.\end{equation}
Using \eq{fourier} to convert this result to Fourier space, we find 
\begin{align} \label{eq:rotation}
 \widetilde X_\bfl &= \intm\, D^\text{Rotation}_{(\bfl,\bfm)}\,X_\bfm
 \nn 
  D^\text{Rotation}_{(\bfl,\bfm)} &= (2\pi)^2 \delta^2(\bfl-\bfm-\cP)\, R_{(-\bfl,\bfm)}
   \nn & \quad \times
  \exp\Big[ 2\lambda \intk\, \alpha_\bfk \Big] 
\,.\end{align}
Here $\lambda$ is the generator of the rotation between $E$ and $B$ modes,
\begin{align}
  \lambda = 
  \begin{pmatrix}
   0 & 0 & 0 \\
   0 & 0 & 1 \\
   0 & -1 & 0
 \end{pmatrix}
\,.\end{align}
Note that although the rotation does not affect temperature, it will affect correlations such as $\langle TE \rangle$ and $\langle TB \rangle$.  
It is straightforward to derive the Feynman rule for the rotated CMB fields from \eq{rotation}, which is shown in \fig{rotation}. Here the power spectrum of rotations $C^{\alpha\alpha}$ is represented by a wiggly line, in analogy to $C^{\phi \phi}$ for the gravitational lensing case. One simplification compared to lensing is that there is no dependence on the momentum of the rotation field. 

It has been shown in Ref.~\cite{2009PhRvD..79l3009Y, 2009PhRvD..80b3510G} that lensing and rotation have orthogonal filters $f^{(\phi,0)}$ and $f^{(\alpha,0)}$, and hence both can be reconstructed using the quadratic estimator formalism without biasing each other's estimate.
However, in contrast to lensing, rotation breaks parity and hence generates  $\widetilde C^{EB}_\ell$ and $\widetilde C^{TB}_{\ell}$. One consequence of this is that the Gaussian noise term $N^{xx',yy'(0)}_L$ for the estimator involving an odd number of observed $B$ fields (for example, $xx'yy'=$ \corr E E E B\ and $xx'yy'=$ \corr T T E B) no longer vanishes, because parity is broken.\\

\subsection{Patchy Reionization}

Reionization marks the birth of the first luminous objects after decoupling. At this stage, the vast majority of Hydrogen becomes ionized due to gravitational non-linearities. When and how this process occurred is at present not well constrained. Observational constraints  from  Lyman-$\alpha$ forest absorption spectra of quasars suggest that reionization ended by redshift $z\approx 6$~\cite{2006AJ....132..117F}. Constraints from the large scale ``reionization bump" imprinted in the CMB polarization \est E E\ and \est T E\ spectra provide a mean redshift of reionization $z=10.5 \pm 1.1$~\cite{2013ApJS..208...19H}.
In addition to constraining the epoch of reionization, the reionization history provides information about the formation of early galaxies. The reionization process is likely to have occurred in a patchy manner, with some regions ionizing early on and other regions remaining neutral until the end of reionization.

Patchy reionization produces several secondary anisotropies in the CMB. Specifically, the patchy nature of reionization results in a direction-dependent Thomson scattering optical depth, $\tau(\bf{n})$. Such optical depth fluctuations modulate CMB fields by suppressing the primordial anisotropies with a factor of $e^{-\tau(\bf{n})}$, correlating different spherical harmonics.  Using tildes to denote the observed CMB components, the effect of patchy reionization is described by
\begin{align} \label{eq:dpatchy}
 \widetilde X_\bfl &= \intm\, D^\text{Reionization}_{(\bfl,\bfm)}\,X_\bfm
 \nn 
  D^\text{Reionization}_{(\bfl,\bfm)} &= (2\pi)^2 \delta^2(\bfl-\bfm-\cP)\, R_{(-\bfl,\bfm)}
   \nn & \quad \times
  \exp\Big[ - \intk\, \tau_\bfk \Big] 
\,.\end{align}
The corresponding Feynman rule is shown in \fig{rotation}.

The filters for patchy reionization and lensing have a strong overlap~\cite{2003PhRvD..67d3004H} , and therefore one can not independently reconstruct $\tau$ and the lensing field $\phi$ from the CMB.
Since the patchy reionization signal is expected to be a few orders of magnitude smaller than the lensing signal,  the estimate of lensing is hardly affected by patchy reionization. However, to extract patchy reionization, one will have to modify the quadratic estimator to avoid the effect of lensing bias.

\begin{figure}[t]
\begin{tabular}{cl}
\raisebox{-0.5\height}{
\psfrag{labelone}[r][r]{$\bfl,x$}
\psfrag{labeltwo}[c][c]{$\bfm,y$}
\psfrag{labelthree}[c][c]{$\bfk_1$}
\psfrag{labelfour}[c][c]{$\bfk_2$}
\psfrag{labelfive}[c][c]{$\bfk_n$}
\includegraphics[width=5cm]{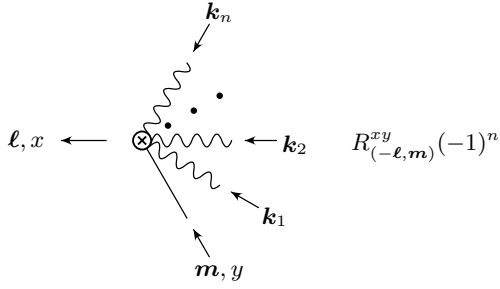}} & $R_{(-\bfl,\bfm)}^{xy} (-1)^n$ \\
\end{tabular}
\caption{Feynman rule for calculating the effect of patchy reionization on the CMB, described by the induced optical depth $\tau$ (wiggly lines). Here $n$ is the number of wiggly lines of the vertex.}
\label{fig:reionization}
\end{figure}

\section{Quadratic estimator for Lensing}\label{sec:quadratic}

\subsection{Quadratic Estimator and Noise}\label{subsec:noise}

Lensing breaks the statistical isotropy of the CMB, correlating the CMB modes as described by \eq{2point}.
This isotropy breaking can be exploited to reconstruct the lensing field from the CMB. 
A quadratic estimator for the gravitational lensing of the CMB can be written generically as
\begin{align} \label{eq:estimator}
\widehat \phi^{xy}_\bfL= \frac{A^{xy}_L}{L^2} \intl\, F^{xy}_{(\bfl,\bfL-\bfl)}
\overline{x}_\bfl \overline{y}_{\bfL-\bfl}  \,,
\end{align}
where $\overline{x}_\bfl$ and $\overline{y}_\bfl$ are the observed CMB modes, including lensing and experimental noise.  The CMB modes $x$ and $y$ can be either $T$, $E$ or $B$, and the estimators will be referred to as \est T T, \est T E, etc.~for $x=T$, $y=T,E$ etc.
The filter $F^{xy}$ in \eq{estimator} is chosen to provide the ``best'' estimate for $\phi$.

The Hu-Okamoto estimator~\cite{Hu:2001kj} is the quadratic estimator that is unbiased and has minimal variance.
Requiring $\widehat \phi$ to be an unbiased estimator, 
\begin{equation}
 \langle \widehat \phi^{xy}_\bfL \rangle_\text{CMB} = \phi_\bfL
\,,\end{equation}
fixes the normalization
\begin{align} \label{eq:A}
A^{xy}_\bfL=L^2 \Big[ \intl\, F^{xy}_{(\bfl,\bfL-\bfl)} f^{(\phi,0)xy}_{(\bfl, \bfL-\bfl)} \Big]^{-1}
\,.\end{align}
\eq{A} follows directly from using \eq{2point} to evaluate $\langle \widetilde x_\bfl \widetilde y_{\bfL-\bfl} \rangle$ at lowest order in $\phi$ (ignoring the pathological $L=0$ contribution). Note that the experimental noise does not enter here, as it does not bias the estimate (see \subsec{exp_noise}).

The variance of this estimator is given by
\begin{align} \label{eq:variance}
&\Big\langle \langle \widehat \phi_\bfL^{xy} \widehat \phi_{\bfL'}^{xy} \rangle_\text{CMB} -  \langle \widehat \phi_\bfL^{xy} \rangle_\text{CMB} \langle \widehat \phi_{\bfL'}^{xy} \rangle_\text{CMB} \Big\rangle_\text{LSS}
\\ &\quad 
= (2\pi)^2 \delta^2(\bfL \!+\! \bfL') \big[N^{xy \zero}_\bfL\!+\!N^{xy \one}_\bfL\!+\!N^{xy (2,c)}_\bfL \!+\! \ord{\phi^6}\big]
\nonumber\,.\end{align}
$N^{xy(0)}_\bfL$ is referred to as Gaussian noise. The higher order noise terms $N^{xy(n)}_\bfL$ involve the connected part of the CMB trispectrum and are of $\ord{\phi^{2n}}$.  

These same noise terms $N^{xy(n)}$ enter as bias in the reconstruction of the lensing power spectrum,
\begin{align} \label{eq:bias}
 \langle \widehat \phi_\bfL^{xy} \widehat \phi_{\bfL'}^{xy} \rangle_\text{CMB,LSS}
&=  (2\pi)^2 \delta^2(\bfL \!+\! \bfL') \big(C^{\phi\phi}_\bfL \!+\! N^{xy \zero}_\bfL\!+\!N^{xy \one}_\bfL
\nonumber \\ & \quad
+\!N^{xy (2,c)}_\bfL \!+\! N^{xy (2,d)}_\bfL \!+\! \ord{\phi^6}\big)
\,.\end{align}
A key point is that at second order and beyond there are disconnected pieces $N^{xy (2,d)}$, $N^{xy (3,d)}$, etc.~that enter the bias of the lensing power spectrum in \eq{bias}, but not the variance in \eq{variance}.

The filter $F^{xy}$ in \eq{estimator} is chosen to minimize the lowest order Gaussian variance $N^{xy(0)}_\bfL$,
\begin{equation} \label{eq:minvar}
   \frac{\delta}{\delta F^{xy}} N^{xy\zero}_\bfL = 0 \ .
\, \end{equation}
The Gaussian noise, given in \eq{N0}, involves the observed spectra $\overline C^{xy}_\bfl$, which include lensing effects and instrumental noise. \eq{minvar} leads to
\begin{widetext}
\begin{align} 
\hspace{-2ex}
2\frac{\int\! d^2\bfl\, \delta F^{xy}_{(\bfl,\bfL-\bfl)} f^{(\phi,0)xy}_{(\bfl, \bfL-\bfl)}}{\int\! d^2\bfl\, F^{xy}_{(\bfl,\bfL-\bfl)} f^{(\phi,0)xy}_{(\bfl, \bfL-\bfl)}}
=
\frac{\int\! d^2\bfl\, \delta F^{xy}_{(\bfl,\bfL-\bfl)}
\Big[ 2 F^{xy}_{(\bfl,\bfL-\bfl)} 
  {\overline C}^{xx}_\bfl {\overline C}^{yy}_{\bfL-\bfl} \!+\! F^{yx}_{(\bfl,\bfL-\bfl)}  {\overline C}^{xy}_\bfl {\overline C}^{xy}_{\bfL-\bfl} \Big] \!+\! \delta F^{yx}_{(\bfl,\bfL-\bfl)} F^{xy}_{(\bfl,\bfL-\bfl)}  {\overline C}^{xy}_\bfl {\overline C}^{xy}_{\bfL-\bfl}}{
  \int\! d^2\bfl\, F^{xy}_{(\bfl,\bfL-\bfl)}
\Big[  F^{xy}_{(\bfl,\bfL-\bfl)} 
  {\overline C}^{xx}_\bfl {\overline C}^{yy}_{\bfL-\bfl} + F^{yx}_{(\bfl,\bfL-\bfl)}  {\overline C}^{xy}_\bfl {\overline C}^{xy}_{\bfL-\bfl} \Big]}
\,,\end{align}
\end{widetext}
by using $F^{xy}_{(\bfL-\bfl,\bfl)}= F^{yx}_{(\bfl,\bfL-\bfl)}$, as well as $F^{xy}_{(-\bfl,\bfl-\bfL)} = F^{xy}_{(\bfl,\bfL-\bfl)}$ (which holds because the lensing potential $\phi$ is parity even). For $x=y$, this implies,
\begin{align} \label{eq:Fsimple}
    F^{xx}_{(\bfl,\bfL-\bfl)} = \frac{f^{(\phi,0)xx}_{(\bfl, \bfL-\bfl)}}{2{\overline C}^{xx}_\bfl {\overline C}^{xx}_{\bfL-\bfl}}
\,,\end{align}
and for $x\neq y$, 
\begin{align}\label{eq:Ffull}
  F^{xy}_{(\bfl,\bfL-\bfl)} = \frac{{\overline C}^{yy}_\bfl {\overline C}^{xx}_{\bfL-\bfl} \ff{xy}{\phi,0}{(\bfl,\bfL-\bfl)} - {\overline C}^{xy}_\bfl {\overline C}^{xy}_{\bfL-\bfl} \ff{xy}{\phi,0}{(\bfL-\bfl,\bfl)}}{{\overline C}^{xx}_\bfl {\overline C}^{xx}_{\bfL-\bfl} {\overline C}^{yy}_\bfl {\overline C}^{yy}_{\bfL-\bfl} - ({\overline C}^{xy}_\bfl {\overline C}^{xy}_{\bfL-\bfl})^2}
\,.\end{align}
Note that the overall normalization of $F^{xy}$ is arbitrary since it cancels against the normalization of $A^{xy}$ and drops out of the estimator.

\subsection{Calculating Noise Terms}
\label{subsec:calc_noise}

\begin{figure}
\begin{tabular}{ccc}
\raisebox{-0.5\height}{\includegraphics[width=0.15\textwidth]{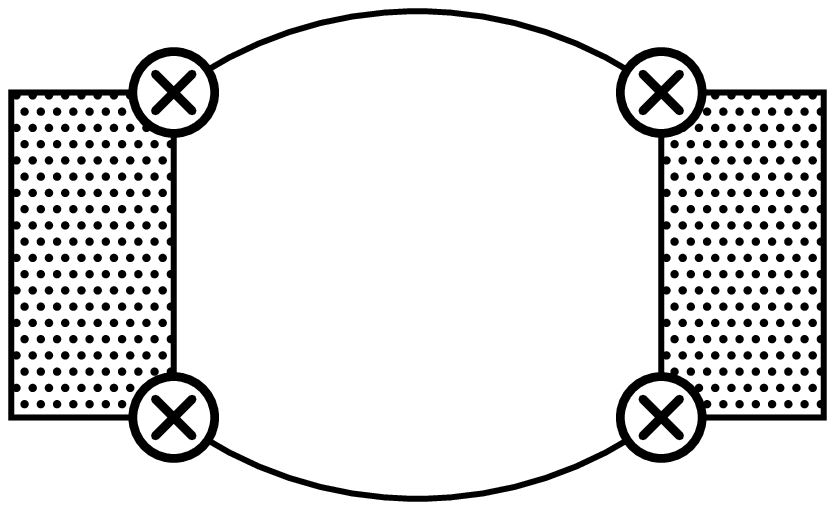}} & \qquad &
\raisebox{-0.5\height}{\includegraphics[width=0.15\textwidth]{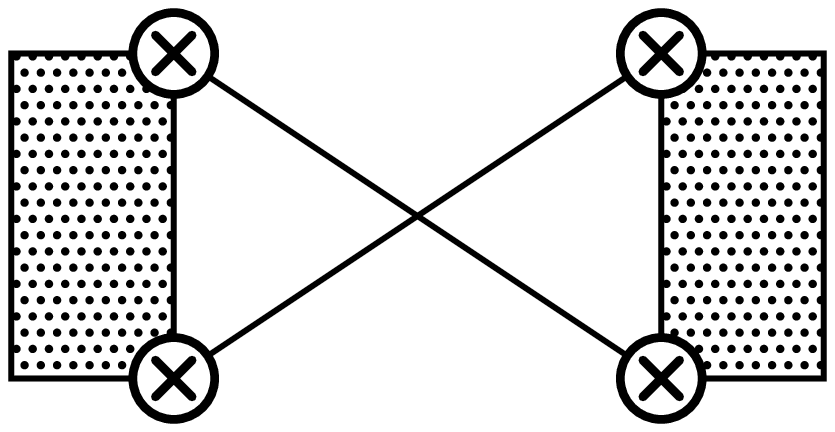}} \\ \ \\[-2ex]
(a) & & (b)
\end{tabular}
\caption{Diagrams describing the lowest-order noise $N^{xy\zero}_\bfL$. The shaded box represents the filter $F^{xy}$ in the quadratic estimator \eq{estimator}.
\label{fig:N0}}
\end{figure}

We will now compute the correlator of two estimators $\langle \widehat \phi^{xy}\ \widehat \phi^{x'y'} \rangle$ up to order $\ord{\phi^4}$, which gives the noise terms $N^{xy,x'y'\zero}_\bfL$, $N^{xy,x'y'\one}_\bfL$, $N^{xy,x'y'(2,c)}_\bfL$ and $N^{xy,x'y'(2,d)}_\bfL$. For $x=x'$ and $y=y'$, this is the noise of the estimator $\widehat \phi^{xy}$ in \eqs{variance}{bias}, but when $x\neq x'$ or $y\neq y'$, it describes the correlation between different estimators.
In fact, we will find in \sec{results} that combining different estimators can help reduce the bias in the determination of $C^{\phi\phi}$.
Using the diagrammatic approach presented here, one can immediately write down expressions for these noise terms and track down the origin of the large contribution to $N^{(2)}$.

The lowest order graphs are shown in \fig{N0}, where each shaded box is one insertion of the estimator $\widehat \phi$. The shaded box represents the filter $F^{xy}$, the normalization constant $A^{xy}$, and the loop integral in \eq{estimator}.
The two fields $\overline x$ and $\overline y$ in the estimator are the lines emerging from the two $\otimes$ symbols at the edge of the shaded box.
Adding up the contributions from the diagrams in \fig{N0}, leads to 
\begin{align} \label{eq:N0}
\hspace{-2ex}N^{xy,x'y'(0)}_\bfL&= \frac{A^{xy}_\bfL A^{x'y'}_\bfL}{L^4} \intl\, F^{xy}_{(\bfl,\bfL-\bfl)}
\Big[  F^{x'y'}_{(-\bfl,\bfl-\bfL)} 
 \nn & \quad \times
  {\overline C}^{xx'}_\bfl {\overline C}^{yy'}_{\bfL-\bfl} + F^{x'y'}_{(\bfl-\bfL,-\bfl)}  {\overline C}^{xy'}_\bfl {\overline C}^{x'y}_{\bfL-\bfl} \Big]
\,,\end{align}
in agreement with Ref.~\cite{Hu:2001kj} for $x=x'$ and $y=y'$. The power spectrum $\overline C$ contains experimental noise (discussed in the next section) and lensing effects. The inclusion of lensing effects is not mandatory, although it is the default for the lowest order noise. It is not the default for higher-order noise terms, though we find that using the lensed power spectrum, as discussed in \subsec{expansion}, improves the convergence of the noise terms.

\subsection{Experimental Noise}
\label{subsec:exp_noise}

The observed power spectra enter in $N^{(0)}$ in \eq{N0}, and thus in the filter $F^{xy}$ in \eqs{Fsimple}{Ffull}, because the quadratic estimator is based on observed CMB modes. 
The observed power spectra
\begin{align} \label{eq:fullC}
{\overline C}^{xy}_\bfl=\widetilde C^{xy}_\bfl+ \Delta^2_{xy} \, e^{\ell (\ell+1)\sigma^2/8\ln2}
\end{align}
include lensing and instrumental noise, where $\sigma$ is the full-width-half-maximum of the experimental beam and  $\Delta_{xy}$ is experimental noise~\cite{Knox:1995dq}. We will assume fully polarized detectors for which $\Delta_{EE}=\Delta_{BB}=\sqrt{2}\Delta_{TT}$, and $\Delta_{xy}=0$ for $x \neq y$. 

Experimental noise does not enter in the higher-order noise terms (except through the filter $F^{xy}$), as we now argue. The observed quantity $\overline x _i =\widetilde x_i + n_i$, where $\widetilde x_i$ is the lensed quantity, and $n_i$ is the experimental noise. The subscript $i$ is shorthand for the CMB mode ($T$, $E$, $B$) as well as the momentum label. Calculations of noise involve the
trispectrum (4-point function),
\begin{align}
 \langle \overline{x}_1 \overline{x}_2 \overline{x}_3 \overline{x}_4 \rangle
 &=
 \langle (\widetilde x_1 + n_1)(\widetilde x_2 + n_2)(\widetilde x_3 + n_3)(\widetilde x_4 + n_4) \rangle
 \nn & =  
  \langle \widetilde x_1 \widetilde x_2 \widetilde x_3 \widetilde x_4 \rangle + (\langle \widetilde x_1 \widetilde x_2 \rangle \langle n_3 n_4 \rangle 
  \nn & \quad 
  + \text{permutations} ) +  \langle n_1 n_2 n_3 n_4 \rangle
  \label{eq:41}
\,.\end{align}
Here we used the fact that the signal $\widetilde x_i$ and the experimental noise $n_i$ are uncorrelated, so that averages such as $\langle \widetilde x_1 \widetilde x_2 n_3 n_4 \rangle$ factor into products $\langle \widetilde x_1 \widetilde x_2 \rangle \langle n_3 n_4 \rangle$, and that averages over odd powers of the experimental noise vanish.
Only the first term $\langle \widetilde x_1 \widetilde x_2 \widetilde x_3 \widetilde x_4 \rangle$ in \eq{41} has corrections beyond $N^{(0)}$, since the experimental noise is Gaussian. The other contributions are pure two-point functions and are therefore fully contained in \eq{N0}.

For correlation functions involving $N\geq 5$ fields, this statement is no longer true, since one can have contributions such as 
$\langle \widetilde x_1 \widetilde x_2 \widetilde x_3 \widetilde x_4 \rangle \langle n_5 n_6 \rangle $, where the first factor has a non-vanishing connected four-point contribution. In this case, there are contributions where the effects of lensing and experimental noise are multiplied rather than simply added. However, we only need terms up to $N=4$ to compute the bias to second order.

\subsection{Higher-Order Noise}
\label{subsec:higher_noise}

\begin{figure}
\begin{tabular}{ccc}
\includegraphics[width=0.175\textwidth]{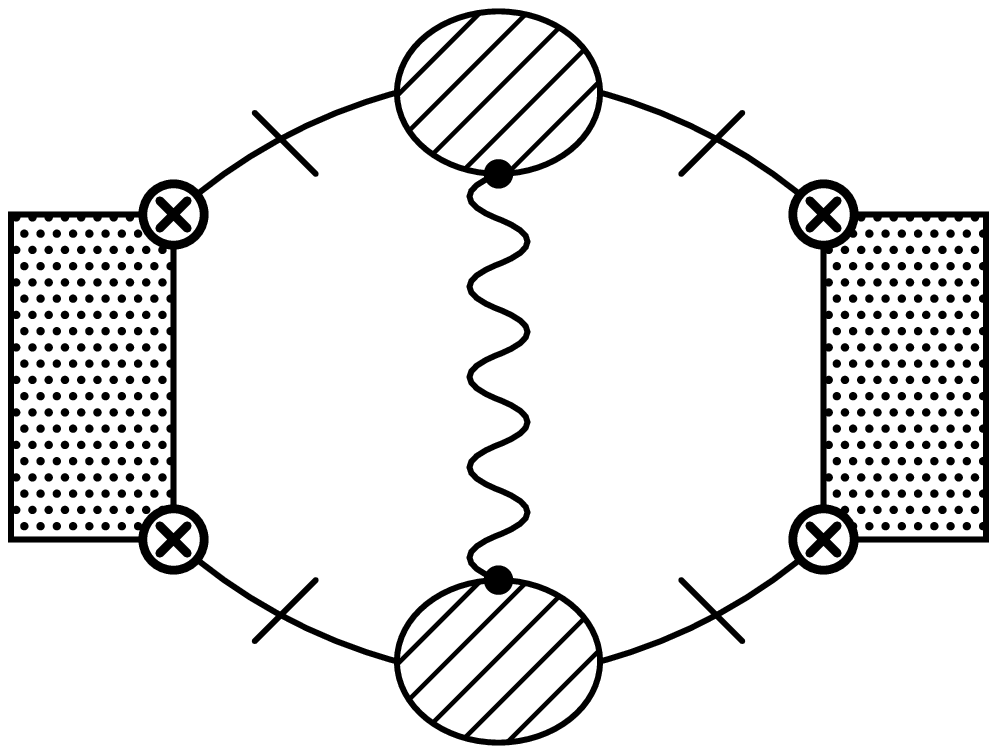} & \qquad &
\raisebox{0.42\height}{\includegraphics[width=0.2\textwidth]{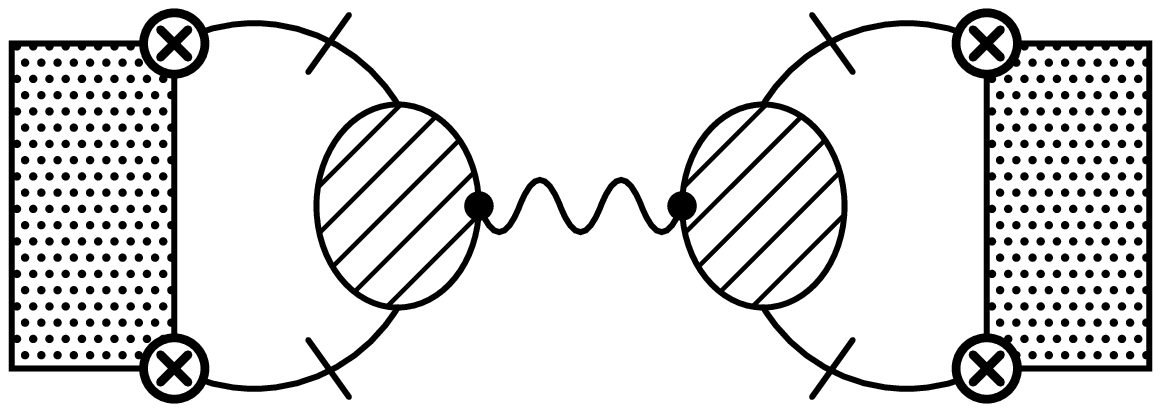}} 
 \\ \ \\[-2ex]
(a) & & (b)
\end{tabular}
\caption{Diagrams contributing to $\langle  \langle \widehat \phi \rangle_{\text{CMB}}  \langle \widehat \phi  \rangle_{\text{CMB}} \rangle_{\text{LSS}}$ at $\ord{\phi^2}$. Diagram (a) and its crossed graph (analogous to \fig{N0}(b)) give $N^\one$. Diagram (b) is disconnected, produces $C^{\phi\phi}$ and does not contribute to the noise $N^{(1)}$.}
\label{fig:N1}
\end{figure}

Diagrams contributing to $\langle \widehat \phi\ \widehat \phi \rangle$  at order $\ord{\phi^2}$ are shown in \fig{N1}.
The shaded blobs in \fig{N1} are the filter $f^{(\phi,0)}$ given in \eq{ff0}.
Diagram \fig{N1}(a) is a connected graph, i.e.~it cannot be disconnected by cutting only $\phi$ lines, and contributes to the noise $N^{(1)}$.
There is also a crossed graph analogous to \fig{N0}(b) which has not been shown. The graph in \fig{N1}(b) is a disconnected diagram, i.e.~it splits into pieces on cutting a $\phi$ line. It does not contribute to the variance in 
\eq{variance}, or to the noise $N^{(1)}$. It does contribute to \eq{bias} where it produces the $C^{\phi\phi}$ term,  since
\begin{align}
C^{\phi \phi}_L\, \bigg( \frac{A^{xy}_\bfL}{L^2}\bigg)^{2}\!\! 
\intl F^{xy}_{(\bfl,\bfL-\bfl)} f^{(\phi,0)xy}_{(\bfl,\bfL-\bfl)}
&\nn 
\times
\intk F^{xy}_{(\bfk,\bfL-\bfk)} f^{(\phi,0)xy}_{(\bfk,\bfL-\bfk)}
&= C^{\phi \phi}_L
\,,\end{align}
which follows trivially from \eq{A}. Note that this is true whether the vertex is the unlensed vertex $f^{(\phi,0)xy}$ or lensed vertex $\widetilde f^{(\phi,0)xy}$.

Using the Feynman rules in \fig{rules}, it is straightforward to obtain the noise $N^{(1)}$ from \fig{N1}(a) and the crossed graph,
\begin{align} \label{eq:N1}
N^{xy,x'y'(1)}_\bfL&= \frac{A^{xy}_\bfL A^{x'y'}_\bfL}{L^4} \intl \frac{d^2 \bfk} {(2\pi)^2}\, C^{\phi \phi}_\bfk F^{xy}_{(\bfl,\bfL-\bfl)} 
\\ & \quad \times
\Big[ F^{x'y'}_{(\bfk-\bfl,\bfl-\bfL-\bfk)} \ff{xx'}{\phi,0}{(\bfl,\bfk-\bfl)} \ff{yy'}{\phi,0}{(\bfL-\bfl,\bfl-\bfL-\bfk)} 
\nonumber \\   &  \qquad 
+ F^{x'y'}_{(\bfl-\bfL-\bfk,\bfk-\bfl)} \ff{xy'}{\phi,0}{(\bfl,\bfk-\bfl)} \ff{yx'}{\phi,0}{(\bfL-\bfl,\bfl-\bfL-\bfk)} \Big] 
\,.\nonumber\end{align}
This noise contribution was first calculated by Kesden et al.~\cite{Cooray:2002py,Kesden:2003cc} for $xy=x'y'$. Their expressions differ from ours for $x \neq y$ because they  symmetrized their filter $F^{xy}_{\bfl,\bfL-\bfl} = F^{yx}_{\bfL-\bfl,\bfl}$.

\begin{figure}
\begin{tabular}{cc}
\raisebox{-1\height}{\includegraphics[width=0.17\textwidth]{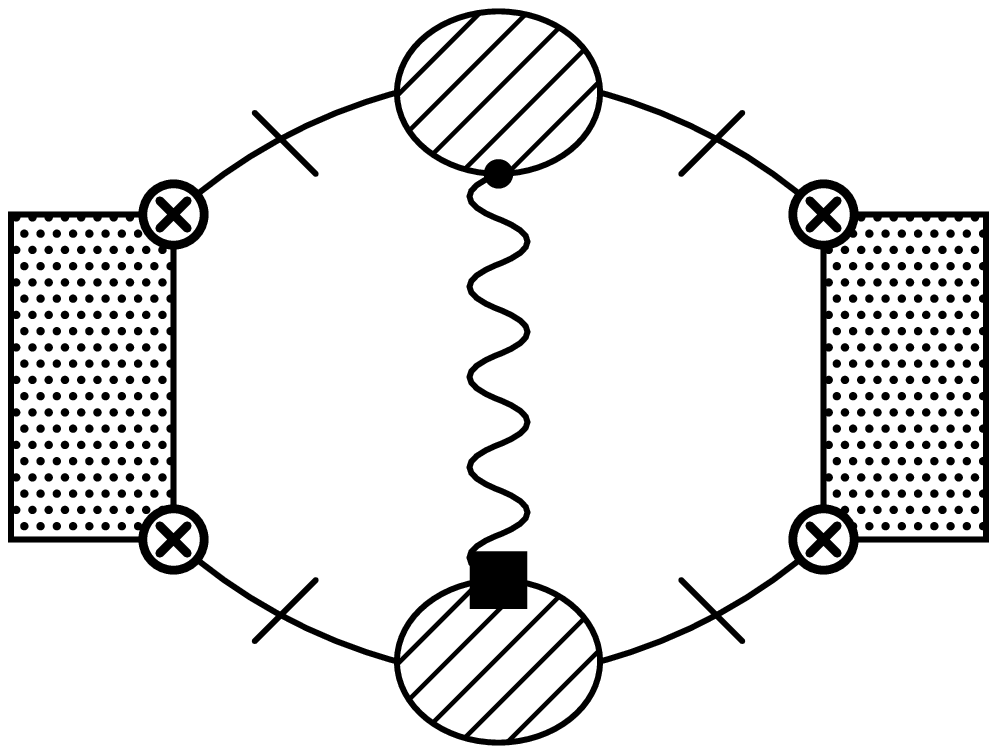}} \quad & \quad
\includegraphics[width=0.17\textwidth,angle=180]{fd31a} \\
(a) \quad & \quad (b) \\[2ex]
\includegraphics[width=0.17\textwidth]{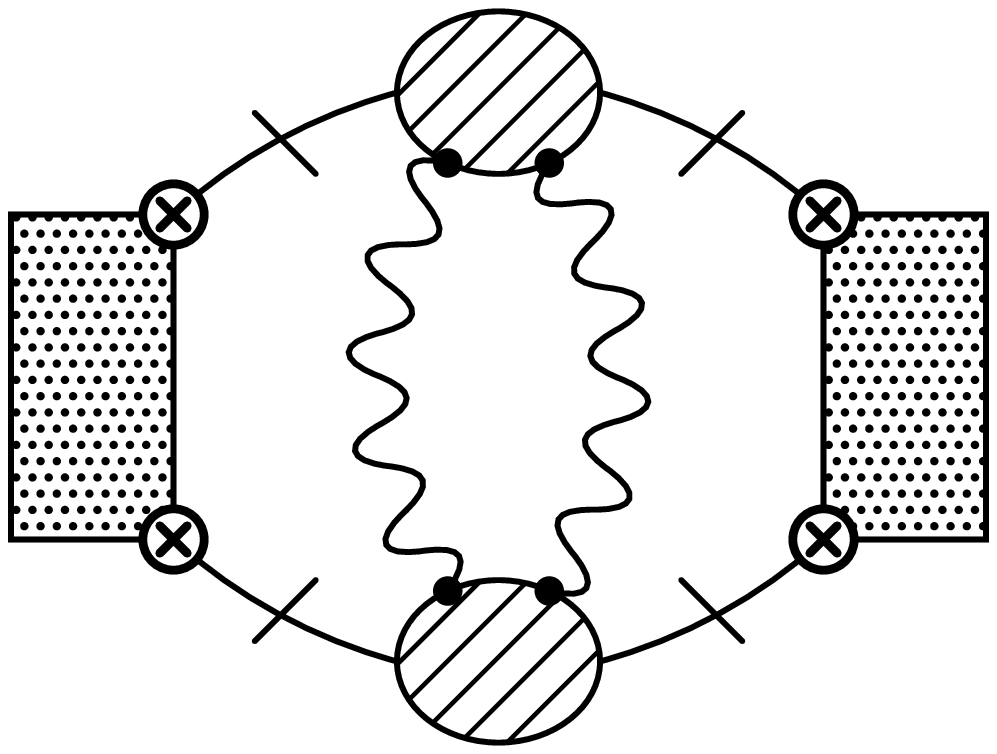} \quad & \quad
\includegraphics[width=0.17\textwidth]{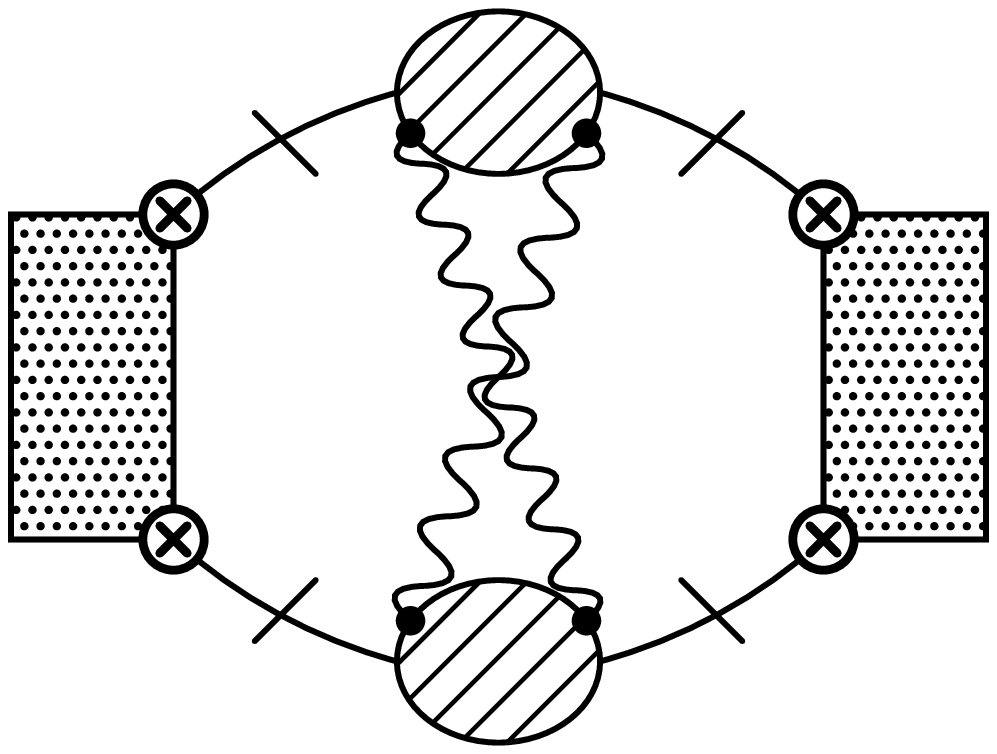} \\
(c) \quad & \quad (d)
\end{tabular}
\caption{Diagrams contributing to $N^{(2,c)}$. The shaded blob is the filter $f^{(\phi\phi,0)}$. The crossed graphs (analogous to \fig{N0}(b)) are not shown.}
\label{fig:N2}
\end{figure}

\begin{figure}
\begin{tabular}{ccc}
\raisebox{-1\height}{\includegraphics[width=0.2\textwidth]{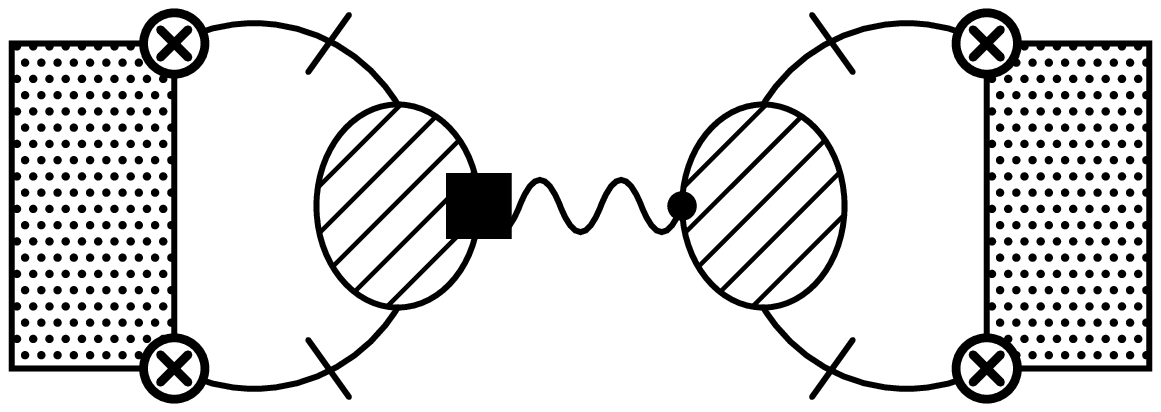}} \quad & \quad  
\raisebox{-0.5\height}{\includegraphics[width=0.2\textwidth,angle=180]{fd38a}} \\
(a) \quad & \quad (b) \\
\raisebox{-0.5\height}{\includegraphics[width=0.2\textwidth]{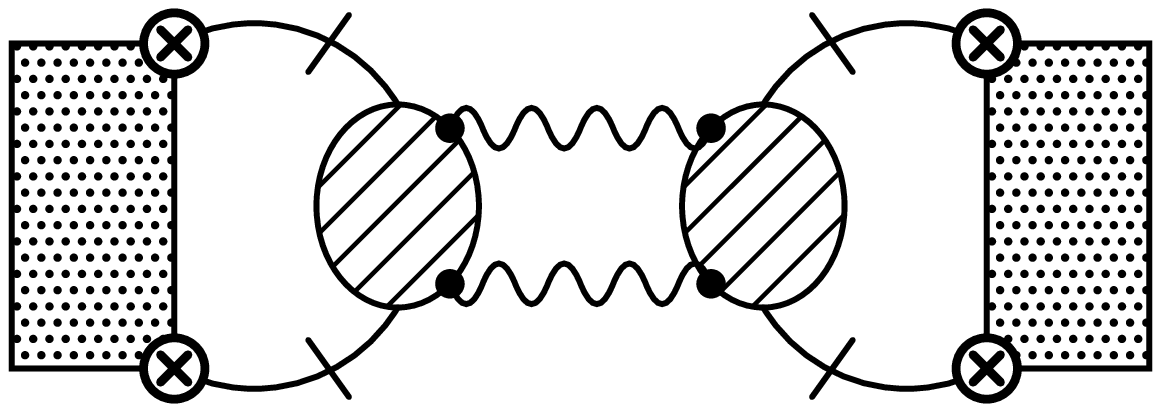}} \quad & \quad
\raisebox{-0.5\height}{\includegraphics[width=0.2\textwidth]{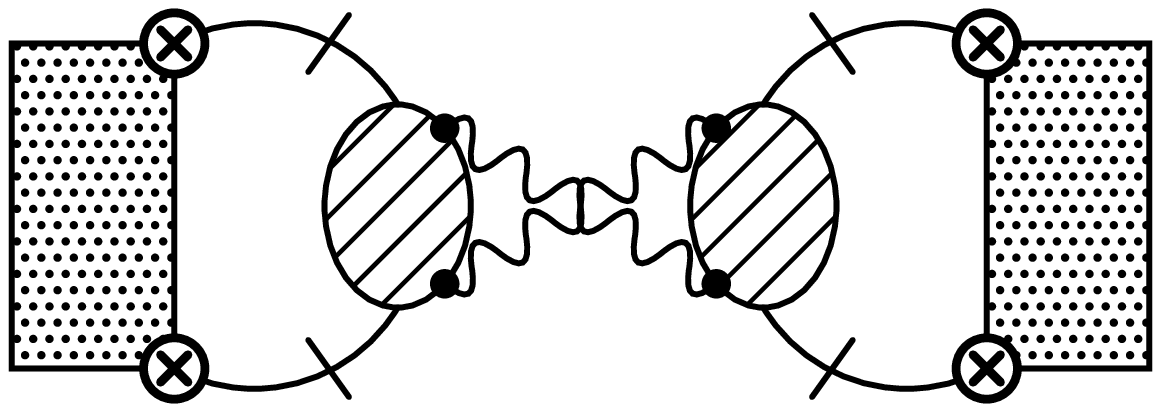}} \\
(c) \quad & \quad (d)
\end{tabular}
\caption{Diagrams contributing to $N^{(2,d)}$. In graphs (a) and (b), one of the blobs is $f^{(\phi,0)}$ and the other (with the black square) is $f^{(\phi,1)}$.}
\label{fig:bias}
\end{figure}

Diagrams contributing to the second order noise $N^{(2)}$ are shown in \figs{N2}{bias}. Again, the crossed graphs for \fig{N2} have not been shown.
The graphs in \fig{N2}(a), (b) and \fig{bias}(a), (b) have one  $f^{(\phi,0)}$ ``blob" and one higher order $f^{(\phi,1)}$ ``blob" marked with a black square.  The shaded blobs in the graphs shown in \fig{N2}(c), (d) and \fig{bias}(c), (d) involve the filter $f^{(\phi\phi,0)}$ in \eq{fphiphi0}, which has two $\phi$ lines.
The connected piece $N^{(2,c)}$, given in \fig{N2}, contributes to both the 
variance and the bias. The disconnected one $N^{(2,d)}$ given in \fig{bias} only contributes to the bias in \eq{bias}, but not to the variance in \eq{variance}. 
The resulting expressions for $N^{(2,c)}$ and $N^{(2,d)}$ are 
\begin{widetext}
\vspace{-3ex}
\begin{align} \label{eq:N2}
N^{xy,x'y'(2,c)}_\bfL&= 
 \frac{A^{xy}_\bfL A^{x'y'}_\bfL}{L^4} \intl \frac{d^2 \bfk}{(2\pi)^2}\, C^{\phi \phi}_\bfk F^{xy}_{(\bfl,\bfL-\bfl)} 
\Big\{ F^{x'y'}_{(\bfk-\bfl,\bfl-\bfL-\bfk)} \big[\ff{xx'}{\phi,0}{(\bfl,\bfk-\bfl)} \ff{yy'}{\phi,1}{(\bfL-\bfl,\bfl-\bfL-\bfk)} +  \ff{xx'}{\phi,1}{(\bfl,\bfk-\bfl)} \ff{yy'}{\phi,0}{(\bfL-\bfl,\bfl-\bfL-\bfk)}\big]
\nn  & \quad
+ F^{x'y'}_{(\bfl-\bfL-\bfk,\bfk-\bfl)} \big[\ff{xy'}{\phi,0}{(\bfl,\bfk-\bfl)} \ff{yx'}{\phi,1}{(\bfL-\bfl,\bfl-\bfL-\bfk)} + \ff{xy'}{\phi,1}{(\bfl,\bfk-\bfl)} \ff{yx'}{\phi,0}{(\bfL-\bfl,\bfl-\bfL-\bfk)}\big]
+ 2 \int\! \frac{d^2 \bfm}{(2\pi)^2}\, C^{\phi\phi}_\bfm \big[ F^{x'y'}_{(\bfk-\bfl+\bfm,\bfl-\bfL-\bfk-\bfm)}
\nn  & \quad \times
 f^{(\phi\phi,0)xx'}_{(\bfl,\bfk-\bfl+\bfm,\bfm)} f^{(\phi\phi,0)yy'}_{(\bfL-\bfl,\bfl-\bfL-\bfk-\bfm,-\bfm)} +
F^{x'y'}_{(\bfl-\bfL-\bfk-\bfm,\bfk-\bfl+\bfm)} f^{(\phi\phi,0)xy'}_{(\bfl,\bfk-\bfl+\bfm,\bfm)} f^{(\phi\phi,0)yx'}_{(\bfL-\bfl,\bfl-\bfL-\bfk-\bfm,-\bfm)}
\big] \Big\} 
\,,\nn 
N^{xy,x'y'(2,d)}_\bfL&= 
 \frac{A^{xy}_\bfL}{L^2}\, C^{\phi\phi}_\bfL \intl F^{xy}_{(\bfl,\bfL-\bfl)} f^{(\phi,1)xy}_{(\bfl,\bfL-\bfl)} +
\frac{A^{x'y'}_\bfL}{L^2}\, C^{\phi\phi}_\bfL \intl F^{x'y'}_{(\bfl,\bfL-\bfl)} f^{(\phi,1)x'y'}_{(\bfl,\bfL-\bfl)} 
\nn & \quad +
2\frac{A^{xy}_\bfL A^{x'y'}_\bfL}{L^4}
\int\! \frac{d^2 \bfl}{(2\pi)^2} \frac{d^2 \bfk}{(2\pi)^2} \frac{d^2 \bfm}{(2\pi)^2}\,
 C^{\phi\phi}_\bfm C^{\phi\phi}_{\bfL-\bfm} 
F^{xy}_{(\bfl,\bfL-\bfl)} F^{x'y'}_{(-\bfk,\bfk-\bfL)} f^{(\phi\phi,0)xy}_{(\bfl,\bfL-\bfl,\bfm)}
 f^{(\phi\phi,0)x'y'}_{(-\bfk,\bfk-\bfL,-\bfm)}\Big\}
\,.\end{align}
\end{widetext}

\begin{figure*}[t] 
\includegraphics[bb=43 310 462 693,width=16cm]{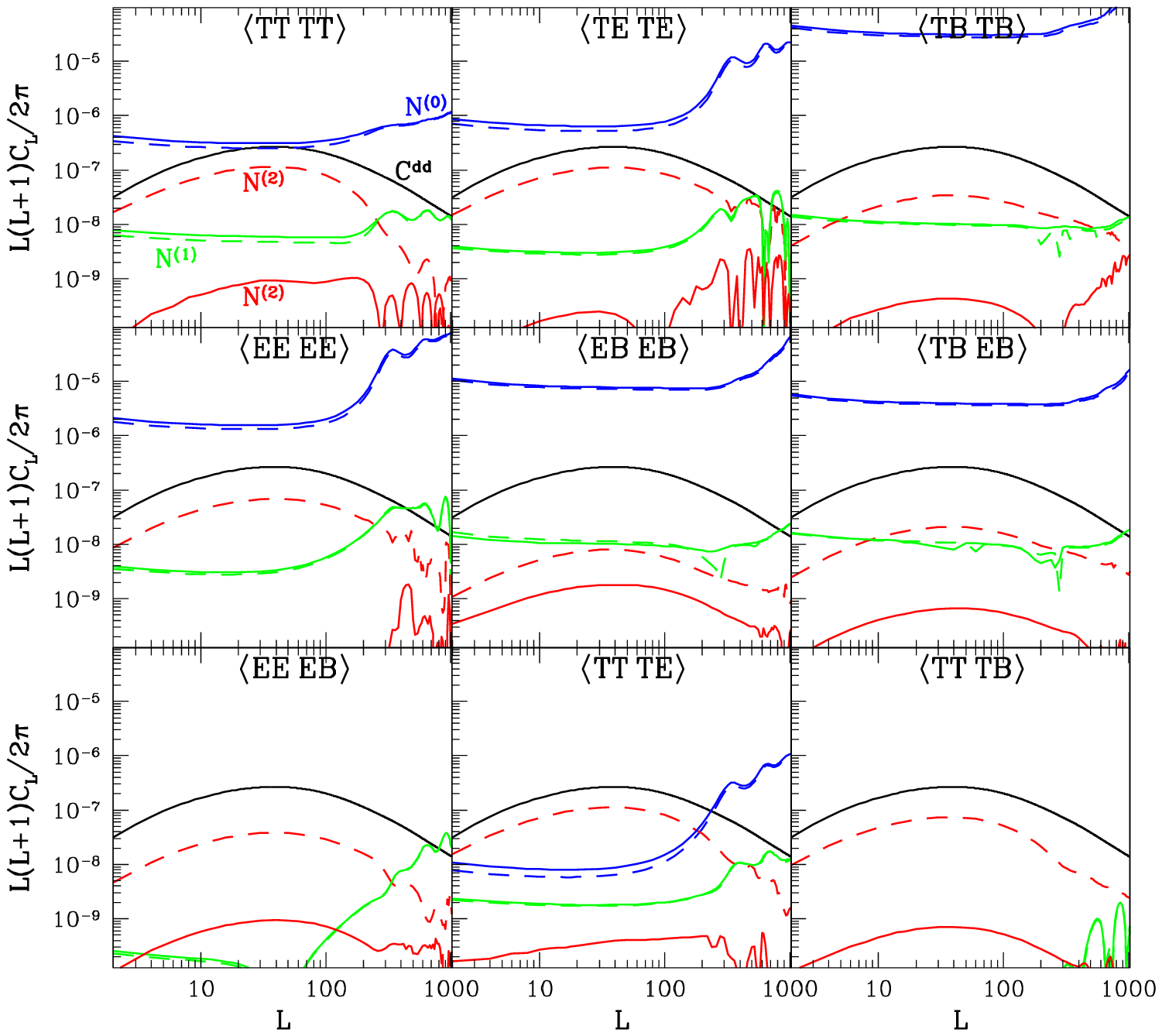}
\caption{Lensing estimator noise $N^{xy,x'y'(n)}_L$ as a function of $L$ for a \planck-like experiment with $\Delta_{EE}=56 \mu$K-arcmin and $\sigma=7$ arcmin. Shown are the theoretical lensing deflection power spectrum $C^{dd}_L = L^2 C^{\phi\phi}_L$ (black curve), the Gaussian noise $N^{(0)}_L$ (blue curves), $N^{(1)}_L$ (green curves) and $N^{(2)}_L$ (red curves). Solid lines represent the noise calculated with our ``lensed counting", and dashed-lines use the  ``unlensed counting". The results for the various estimators $xy,x'y'$ are shown in separate panels. The noise $N^{(2)}_L$ is negative for the unlensed counting while it changes sign for the lensed-counting $N^{(2)}_L$. For clarity we only show the magnitude here, but the sign is shown in Figs.~\ref{fig:noise_lensed} and~\ref{fig:N_contributions}. }
\label{fig:N_exp1}
\end{figure*}

\subsection{Large $N^{(2)}$ Contribution}\label{subsec:n2}

We can now address the large $N^{(2)}$ contribution to the bias in \eq{bias}, that was previously observed in Refs.~\cite{Hanson:2010rp,Anderes:2013jw}. For the lensing estimator method to work, it is crucial that the $\phi$ power series expansion is well-behaved such that higher-order noise terms converge  $N^{(0)} \gg N^{(1)} \gg N^{(2)} \gg \dots$. The large size of $N^{(2)}$ is a problem because it indicates that higher-order corrections may not be under control. We will show that this is not the case: the expansion is well-behaved, and the second order noise can be reliably calculated. 

A first clue to the solution is provided by the fact that the large contribution arises from the disconnected contribution $N^{(2,d)}$, whereas the connected contribution $N^{(2,c)}$ is as small as one might expect. It is insightful to first compare the connected and disconnected contribution at $\ord{\phi^2}$, which are shown in \fig{N1}. The connected graph in \fig{N1}(a) gives the noise $N^{(1)}$, and the disconnected graph in \fig{N1}(b) gives the lensing power spectrum $C^{\phi\phi}_\bfL$. 
Both terms are formally order $\ord{\phi^2}$, but one can see from \fig{N_exp1}, where these quantities are plotted numerically, that the power spectrum is  larger than the noise $N^{(1)}$. As is clear from \eq{N1}, the calculation of $N^{(1)}$ involves two loop integrals over unconstrained internal momenta. By contrast, \fig{N1}(b) gives exactly $C^{\phi\phi}_L$ and has no such integrals. It is well known in quantum field theory that loop integrals are suppressed. Here we have two-dimensional loop integrals which, in field theory, would bring an extra $1/(4\pi)$ loop suppression factor. The loop suppression factor is well understood in field theory, but there the calculations are much simpler because they involve propagators of the form $1/(\bfl^2+M^2)$. In the lensing problem, the propagators are the spectra and have a more complicated form, but the same loop suppression is at work.

At order $\ord{\phi^4}$, the bias $N^{(2)}$ gets contributions from the connected graphs in \fig{N2} and the disconnected graphs in \fig{bias}. The graphs in \fig{bias}(a) and (b) are the $\ord{\phi^4}$ analog of \fig{N1}(b). They involve one less loop integral than the graphs in \fig{N2} and \fig{bias}(c) and (d), as can be seen from comparing the first line for $N^{(2,d)}$ in \eq{N2} with the other terms. Indeed, numerically, \fig{bias}(a) and (b) are responsible for the large value of $N^{(2)}$.
We thus have two series of terms that contribute to the bias, which are schematically of the form
$a_0 + a_2 \phi^2 + a_4 \phi^4 + \ldots$ and $b_0 + b_2 \phi^2 + b_4 \phi^4 + \ldots$. The $a_i$ series has less loop integrals than the corresponding terms in the $b_i$ series, so that $a_i \gg b_i$.
The $\phi$ expansion is well-behaved for each series, and one expects the $a_6 \phi^6 $ term to be smaller than the  $a_4 \phi^4$ term, and the $b_6 \phi^6$ term to be smaller than the $b_4 \phi^4$ term. The reason there appears to be a lack of convergence is because the $a$ series first starts at $\ord{\phi^4}$ from the graphs in \fig{bias}(a) and (b), as there are no $\ord{\phi^0}$ or $\ord{\phi^2}$ contributions. (The $\ord{\phi^2}$ graph in \fig{N1}(b) yields $C^{\phi\phi}$, and does not contribute to the $a_2$ term in the noise.)

We can take care of this larger $a$-series by reorganizing the perturbation expansion using lensed spectra, as discussed in \subsec{expansion}. This reorganization absorbs most of the higher order terms from \fig{bias}(a) and (b) into the lower order \fig{N1}(b). However, because we also changed the filter $F$ of the quadratic estimator accordingly, \fig{N1}(b) still exactly gives $C^{\phi\phi}_L$.

\section{Numerical Results}\label{sec:results}

\begin{figure*}[] 
\includegraphics[bb=43 310 462 693,width=16cm]{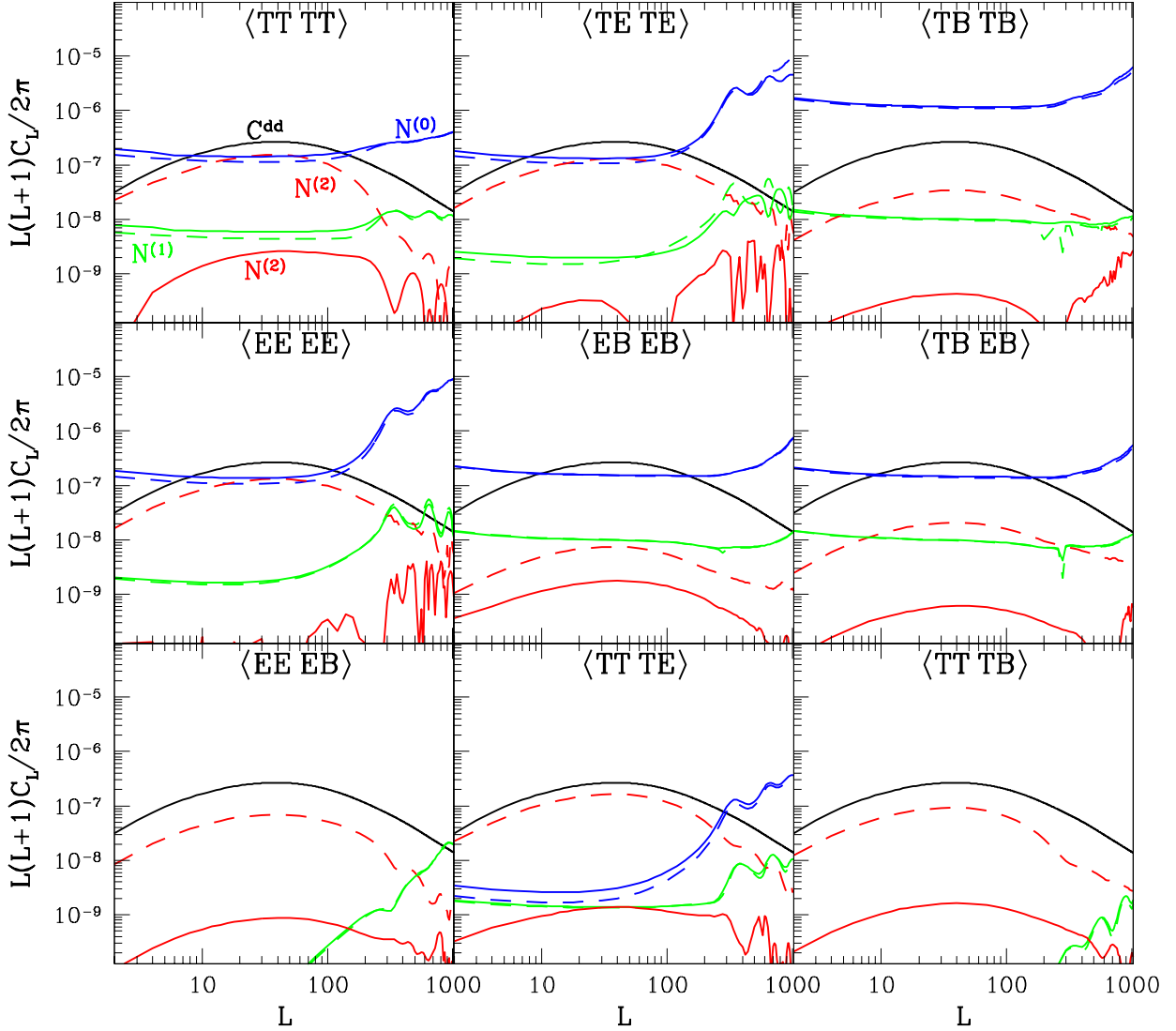}
\caption{Same as Figure~\ref{fig:N_exp1} but for an experiment with $\Delta_{EE}=10 \mu$K-arcmin and $\sigma=8$ arcmin.}
\label{fig:N_exp2}
\end{figure*}

In Figure~\ref{fig:N_exp1} and~\ref{fig:N_exp2} the lensing noise $N^{xy,x'y'(n)}_L$ is shown as a function of multipole moment $L$, which enters as bias in the reconstruction of the lensing power spectrum in \eq{bias}. We present results for various CMB channels \corr{$x$}{$y$}{$x'$}{$y'$}, going beyond previous studies by considering the cases $x \neq x'$ and/or $y \neq y'$. These noise terms are shown for both the default unlensed counting (dashed curves), as well as the lensed counting (solid curves) that we introduce in \subsec{expansion}.
We have considered two experimental setups:
\begin{itemize}
\item \fig{N_exp1}: a \planck-like experiment with polarization instrumental noise $\Delta_{EE}\equiv\Delta_{EE}=\Delta_{BB}=56 \mu$K-arcmin and Gaussian beam full-width-half-maximum (FWHM), $\sigma=7$ arcmin.
\item \fig{N_exp2}: An experiment representative of current generation balloon and ground-based experiments with instrumental noise $\Delta_{EE}=10 \mu$K-arcmin and Gaussian beam FWHM, $\sigma=8$ arcmin.
\end{itemize}

The unlensed counting shows the so called ``$N^{(2)}$-bias", i.e. on large scales ($\bfl <$ few hundred), $N^{(2)}_L$ becomes larger than $N^{(0)}_L$ and $N^{(1)}_L$. The $N^{(2)}$-bias in the \corr T T T T\ channel was first discovered and discussed in Ref.~\cite{Hanson:2010rp}. 
In Ref.~\cite{Anderes:2013jw}, $N^{(2)}_L$ for the polarization channels were studied using simulations.  This paper is the first time the exact calculation of CMB polarization channel based $N^{(2)}_L$ has been performed. 
As can be seen from the red dashed curves in \figs{N_exp1}{N_exp2}, the $N^{(2)}$-bias is largest for the \corr T T T T\ estimator and smallest for the \corr E B E B\ estimator. In general, estimators involving temperature maps tend to have a larger $N^{(2)}$-bias compared to the ones involving polarization.

For our lensed-counting (solid curves), the $N^{(2)}_L$-bias is absorbed in the lower order noise terms $N^{(0)}_L$ and $N^{(1)}_L$. Note that this method not only reshuffles contributions between the noise terms, but also changes the quadratic estimator by modifying the filter $F$. Therefore the sum of the noise terms is not equal in the two cases. The  lensed counting preserves a desired expansion property for all the estimators, $C^{dd}_L\gg N^{(2)}_L$ and $C^{dd}_L \gg N^{(1)}_L$. 
 
\begin{figure*}[] 
\includegraphics[bb=43 310 462 693,width=16cm]{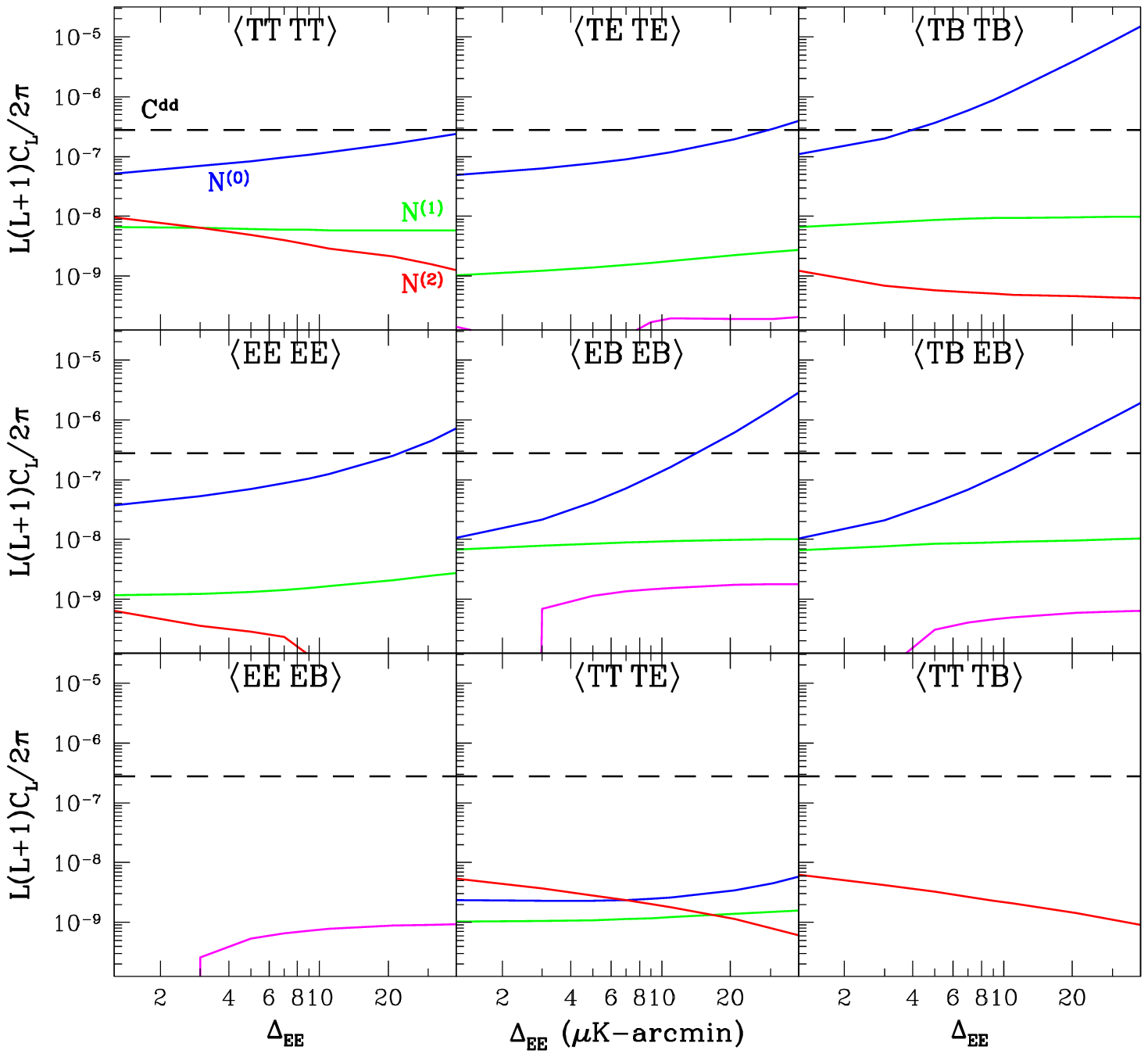}
\caption{Estimator noise $N^{(n)}_{L=40}$ for the lensed counting as a function of instrumental noise $\Delta_{EE}$ with beam FWHM $\sigma=8$ arcmin. 
Shown are the lensing deflection power spectrum $C^{dd}_L$ (black dashed), the Gaussian noise $N^{(0)}_L$ (blue curve), $N^{(1)}_L$ (green curve) and $N^{(2)}_L$ (magenta/red curve). The noise $N^{(2)}$ changes sign, so we use magenta for positive values and show the magnitude of negative values in red.}
\label{fig:noise_lensed}
\end{figure*}

 For the lensed-counting the $N^{xy,x'y'(0)}_L$ noise is the dominant noise term for the estimators with even numbers of $B$ fields (e.g. \corr T T T T\  and \corr T B E B). However the noise $N^{xx'yy'(0)}_L$ for channels involving odd numbers of $B$ fields (e.g.\ \corr E E E B\ and \corr T T E B)  vanishes. This is because the physics responsible for decoupling is parity conserving, and the correlation between parity-even field $T$ or $E$ and parity-odd field $B$ vanishes, $C^{EB}_\ell = C^{TB}_\ell=0$. 
Thus for estimators with an odd number of $B$ fields, the dominant noise comes from the higher order $N^{(n)}_L$ (with $n>0$). Interestingly, for our lensed counting these higher order noises are more than two orders of magnitude smaller than the signal $C^{dd}_L$. As shown in \fig{noise_lensed} and discussed below, this is true for a wide range of experiments. It is worth emphasizing that although $N^{(2)} > N^{(1)}$ for a range of $L$ of \corr E E E B\ and \corr T T T B, this does not pose an inconsistency. As we discussed in \subsec{n2}, there are two types of diagrams that form two separately converging series. This only requires $N^{(2,c)} < N^{(1)}$ and $N^{(2,d)} < C^{dd}$, which are both satisfied.

In \fig{noise_lensed}, we show estimator noise $N^{(n)}_{L=40}$ at $L=40$ for the lensed counting as a function of the instrumental noise $\Delta_{EE}$.  As may be expected from \subsec{exp_noise}, $N^{(0)}_L$ is the most sensitive to experimental noise because it directly gets added to it. For $N^{(1)}_L$ and $N^{(2)}_L$ the experimental noise only appears indirectly through the filter $F$ of the quadratic estimator. 
For the range of experimental noise considered ($\Delta_{EE}=1-60~\mu$K-arcmin), the estimator with lensed-counting will not be biased by $N^{(1)}_L$ or $N^{(2)}_L$, as these higher order noises are  at least an order of magnitude smaller than the signal $C^{dd}_L$. 
  
\begin{figure*} 
\includegraphics[bb=33 544 337 688,width=12cm]{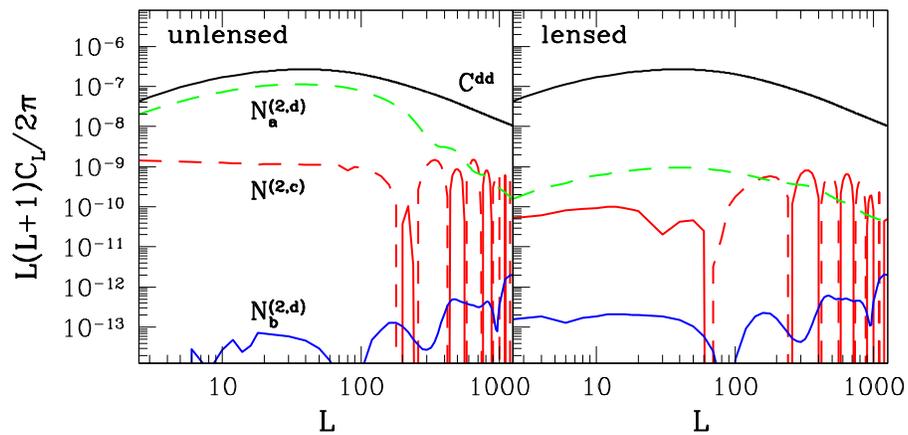}
\caption{Contributions to the lensing estimator noise $N^{TT,TT(2)}_L$ as a function of $L$ for a \planck-like experiment. 
The left panel uses the unlensed counting and the right one our lensed counting.
Shown are the lensing power spectrum $C^{dd}_L$ (solid black), the connected contribution $N^{(2,c)}$ (red) and the disconnected contribution separated into  $N^{(2,d)}_a$ (green) and $N^{(2,d)}_b$ (blue), corresponding to the fourth and fifth line of \eq{N2}. When the contributions to $N^{TT,(2)}$ are positive they are shown as solid lines, and when they are negative, they are shown as dashed lines. }
\label{fig:N_contributions}
\end{figure*}

Figure~\ref{fig:N_contributions} shows three pieces that contribute to $N^{TTTT(2)}_L$, the connected piece $N^{(2,c)}_L$ (dashed) and the disconnected piece split up in the contribution from the fourth line $N^{(2,d)}_a$ (solid) and fifth line $N^{(2,d)}_b$ (dot-dashed) in \eq{N2}. The largest contribution on scales $L\lesssim 200$ comes from $N^{(2,d)}_a$, because it has one fewer loop integral than other pieces (see \subsec{n2} for more details). 
Note that there is never a large $N^{(2)}$ contribution to the variance of the lensing, since only the connected contributions enters in \eq{variance}.
 
\section{Summary and Discussion}\label{sec:summary}
 
We have calculated noise properties of Hu-Okamoto-based quadratic estimators of CMB lensing, using a Feynman diagram approach~\cite{2014arXiv1403.2386J}. This method allowed us, for the first time, to obtain analytical expressions for the higher order noise (up to $\ord{\phi^{4}}$) of lensing estimators based on any combination of CMB temperature and polarization channels. Previous analytical calculations were limited to the temperature channel (\corr T T T T) at this order. We have also discussed how to extend this calculation to other distorting fields like patchy reionization and cosmic rotation, deriving the relevant Feynman rules.
 
 Using this approach, it was straightforward to identify the origin of the (supposed) poor convergence of higher order noise terms. The previously noted large $\ord{\phi^4}$ term in the second order noise $N^{(2)}_L$ has been identified to come from a particular class of diagrams. By reorganizing the $\phi$ expansion, we significantly reduced the effect of higher order noise terms. We have shown results for the estimator noise up to $\ord{\phi^{4}}$ for 9 channels and for a wide range of experimental setups. From this computation we conclude that, using our re-arranged counting, the estimator is well behaved for all the channels. We also note that estimators with an odd number of $B$-fields have a very small noise.

With more precision CMB polarization experiments on the way, it is extremely important to understand and improve the statistical techniques used to extract the lensing information from the data. High precision lensing maps open the door to constraining several fundamental cosmological parameters, the sum of the neutrino masses and the properties of dark energy.  Perhaps most importantly, characterizing lensing opens up the possibility of ``delensing" which enhances sensitivity to measure inflationary $B$-modes induced by tensor perturbations. This is important for $L \gtrsim 100$ when the lensing contribution is no longer small compared to primordial $B$-modes.

\acknowledgments{APSY would like to thank Matias Zaldarriaga for discussions at an early stage of this project. EJ and AM are supported by the U.S.~Department of Energy through DOE grant DE-SC0009919. WJW is supported by Marie Curie Fellowship PIIF-GA-2012-328913. APSY acknowledges support from the Ax Center for Experimental Cosmology at UC San Diego.}

\bibliography{lensing}

\begin{thebibliography}{65}
\expandafter\ifx\csname natexlab\endcsname\relax\def\natexlab#1{#1}\fi
\expandafter\ifx\csname bibnamefont\endcsname\relax
  \def\bibnamefont#1{#1}\fi
\expandafter\ifx\csname bibfnamefont\endcsname\relax
  \def\bibfnamefont#1{#1}\fi
\expandafter\ifx\csname citenamefont\endcsname\relax
  \def\citenamefont#1{#1}\fi
\expandafter\ifx\csname url\endcsname\relax
  \def\url#1{\texttt{#1}}\fi
\expandafter\ifx\csname urlprefix\endcsname\relax\def\urlprefix{URL }\fi
\providecommand{\bibinfo}[2]{#2}
\providecommand{\eprint}[2][]{\url{#2}}

\bibitem[{\citenamefont{Seljak and Zaldarriaga}(1997)}]{Seljak:1996gy}
\bibinfo{author}{\bibfnamefont{U.}~\bibnamefont{Seljak}} \bibnamefont{and}
  \bibinfo{author}{\bibfnamefont{M.}~\bibnamefont{Zaldarriaga}},
  \bibinfo{journal}{Phys. Rev. Lett.} \textbf{\bibinfo{volume}{78}},
  \bibinfo{pages}{2054} (\bibinfo{year}{1997}), \eprint{astro-ph/9609169}.

\bibitem[{\citenamefont{Kamionkowski
  et~al.}(1997{\natexlab{a}})\citenamefont{Kamionkowski, Kosowsky, and
  Stebbins}}]{Kamionkowski:1996ks}
\bibinfo{author}{\bibfnamefont{M.}~\bibnamefont{Kamionkowski}},
  \bibinfo{author}{\bibfnamefont{A.}~\bibnamefont{Kosowsky}}, \bibnamefont{and}
  \bibinfo{author}{\bibfnamefont{A.}~\bibnamefont{Stebbins}},
  \bibinfo{journal}{Phys. Rev.} \textbf{\bibinfo{volume}{D55}},
  \bibinfo{pages}{7368} (\bibinfo{year}{1997}{\natexlab{a}}),
  \eprint{astro-ph/9611125}.

\bibitem[{\citenamefont{Kamionkowski
  et~al.}(1997{\natexlab{b}})\citenamefont{Kamionkowski, Kosowsky, and
  Stebbins}}]{Kamionkowski:1996zd}
\bibinfo{author}{\bibfnamefont{M.}~\bibnamefont{Kamionkowski}},
  \bibinfo{author}{\bibfnamefont{A.}~\bibnamefont{Kosowsky}}, \bibnamefont{and}
  \bibinfo{author}{\bibfnamefont{A.}~\bibnamefont{Stebbins}},
  \bibinfo{journal}{Phys. Rev. Lett.} \textbf{\bibinfo{volume}{78}},
  \bibinfo{pages}{2058} (\bibinfo{year}{1997}{\natexlab{b}}),
  \eprint{astro-ph/9609132}.

\bibitem[{\citenamefont{Baumann et~al.}(2009)}]{2009AIPC.1141...10B}
\bibinfo{author}{\bibfnamefont{D.}~\bibnamefont{Baumann}} \bibnamefont{et~al.}
  (\bibinfo{collaboration}{CMBPol Study Team}), \bibinfo{journal}{AIP Conf.
  Proc.} \textbf{\bibinfo{volume}{1141}}, \bibinfo{pages}{10}
  (\bibinfo{year}{2009}), \eprint{arXiv:0811.3919}.

\bibitem[{\citenamefont{Ade et~al.}(2014)}]{Ade:aa}
\bibinfo{author}{\bibfnamefont{P.}~\bibnamefont{Ade}} \bibnamefont{et~al.}
  (\bibinfo{collaboration}{BICEP 2 Collaboration}) (\bibinfo{year}{2014}),
  \eprint{arXiv:1403.3985}.

\bibitem[{\citenamefont{{Zaldarriaga} and
  {Seljak}}(1998)}]{1998PhRvD..58b3003Z}
\bibinfo{author}{\bibfnamefont{M.}~\bibnamefont{{Zaldarriaga}}}
  \bibnamefont{and} \bibinfo{author}{\bibfnamefont{U.}~\bibnamefont{{Seljak}}},
  \bibinfo{journal}{Phys. Rev.} \textbf{\bibinfo{volume}{D58}},
  \bibinfo{eid}{023003} (\bibinfo{year}{1998}), \eprint{astro-ph/9803150}.

\bibitem[{\citenamefont{{Smith} et~al.}(2007)\citenamefont{{Smith}, {Zahn}, and
  {Dor{\'e}}}}]{2007PhRvD..76d3510S}
\bibinfo{author}{\bibfnamefont{K.~M.} \bibnamefont{{Smith}}},
  \bibinfo{author}{\bibfnamefont{O.}~\bibnamefont{{Zahn}}}, \bibnamefont{and}
  \bibinfo{author}{\bibfnamefont{O.}~\bibnamefont{{Dor{\'e}}}},
  \bibinfo{journal}{Phys. Rev.} \textbf{\bibinfo{volume}{D76}},
  \bibinfo{eid}{043510} (\bibinfo{year}{2007}), \eprint{arXiv:0705.3980}.

\bibitem[{\citenamefont{{Hirata} et~al.}(2008)\citenamefont{{Hirata}, {Ho},
  {Padmanabhan}, {Seljak}, and {Bahcall}}}]{2008PhRvD..78d3520H}
\bibinfo{author}{\bibfnamefont{C.~M.} \bibnamefont{{Hirata}}},
  \bibinfo{author}{\bibfnamefont{S.}~\bibnamefont{{Ho}}},
  \bibinfo{author}{\bibfnamefont{N.}~\bibnamefont{{Padmanabhan}}},
  \bibinfo{author}{\bibfnamefont{U.}~\bibnamefont{{Seljak}}}, \bibnamefont{and}
  \bibinfo{author}{\bibfnamefont{N.~A.} \bibnamefont{{Bahcall}}},
  \bibinfo{journal}{Phys. Rev.} \textbf{\bibinfo{volume}{D78}},
  \bibinfo{eid}{043520} (\bibinfo{year}{2008}), \eprint{arXiv:0801.0644}.

\bibitem[{\citenamefont{Das et~al.}(2011)\citenamefont{Das, Sherwin
  et~al.}}]{2011PhRvL.107b1301D}
\bibinfo{author}{\bibfnamefont{S.}~\bibnamefont{Das}},
  \bibinfo{author}{\bibfnamefont{B.~D.} \bibnamefont{Sherwin}},
  \bibnamefont{et~al.}, \bibinfo{journal}{Phys. Rev. Lett.}
  \textbf{\bibinfo{volume}{107}}, \bibinfo{pages}{021301}
  (\bibinfo{year}{2011}), \eprint{arXiv:1103.2124}.

\bibitem[{\citenamefont{Ade et~al.}(2013{\natexlab{a}})}]{Ade:2013tyw}
\bibinfo{author}{\bibfnamefont{P.}~\bibnamefont{Ade}} \bibnamefont{et~al.}
  (\bibinfo{collaboration}{Planck Collaboration})
  (\bibinfo{year}{2013}{\natexlab{a}}), \eprint{arXiv:1303.5077}.

\bibitem[{\citenamefont{Hanson et~al.}(2013)}]{2013PhRvL.111n1301H}
\bibinfo{author}{\bibfnamefont{D.}~\bibnamefont{Hanson}} \bibnamefont{et~al.}
  (\bibinfo{collaboration}{SPTpol Collaboration}), \bibinfo{journal}{Phys. Rev.
  Lett.} \textbf{\bibinfo{volume}{111}}, \bibinfo{pages}{141301}
  (\bibinfo{year}{2013}), \eprint{arXiv:1307.5830}.

\bibitem[{\citenamefont{Ade et~al.}(2013{\natexlab{b}})}]{2013arXiv1312.6646P}
\bibinfo{author}{\bibfnamefont{P.}~\bibnamefont{Ade}} \bibnamefont{et~al.}
  (\bibinfo{collaboration}{POLARBEAR Collaboration})
  (\bibinfo{year}{2013}{\natexlab{b}}), \eprint{arXiv:1312.6646}.

\bibitem[{\citenamefont{Ade et~al.}(2013{\natexlab{c}})}]{2013arXiv1312.6645P}
\bibinfo{author}{\bibfnamefont{P.}~\bibnamefont{Ade}} \bibnamefont{et~al.}
  (\bibinfo{collaboration}{POLARBEAR Collaboration})
  (\bibinfo{year}{2013}{\natexlab{c}}), \eprint{arXiv:1312.6645}.

\bibitem[{\citenamefont{{Bond} et~al.}(1997)\citenamefont{{Bond}, {Efstathiou},
  and {Tegmark}}}]{1997MNRAS.291L..33B}
\bibinfo{author}{\bibfnamefont{J.~R.} \bibnamefont{{Bond}}},
  \bibinfo{author}{\bibfnamefont{G.}~\bibnamefont{{Efstathiou}}},
  \bibnamefont{and}
  \bibinfo{author}{\bibfnamefont{M.}~\bibnamefont{{Tegmark}}},
  \bibinfo{journal}{MNRAS} \textbf{\bibinfo{volume}{291}}, \bibinfo{pages}{L33}
  (\bibinfo{year}{1997}), \eprint{astro-ph/9702100}.

\bibitem[{\citenamefont{{Zaldarriaga} et~al.}(1997)\citenamefont{{Zaldarriaga},
  {Spergel}, and {Seljak}}}]{1997ApJ...488....1Z}
\bibinfo{author}{\bibfnamefont{M.}~\bibnamefont{{Zaldarriaga}}},
  \bibinfo{author}{\bibfnamefont{D.~N.} \bibnamefont{{Spergel}}},
  \bibnamefont{and} \bibinfo{author}{\bibfnamefont{U.}~\bibnamefont{{Seljak}}},
  \bibinfo{journal}{\apj} \textbf{\bibinfo{volume}{488}}, \bibinfo{pages}{1}
  (\bibinfo{year}{1997}), \eprint{astro-ph/9702157}.

\bibitem[{\citenamefont{{Metcalf} and {Silk}}(1998)}]{1998ApJ...492L...1M}
\bibinfo{author}{\bibfnamefont{R.~B.} \bibnamefont{{Metcalf}}}
  \bibnamefont{and} \bibinfo{author}{\bibfnamefont{J.}~\bibnamefont{{Silk}}},
  \bibinfo{journal}{Astrophys. J. Lett.} \textbf{\bibinfo{volume}{492}},
  \bibinfo{pages}{L1} (\bibinfo{year}{1998}), \eprint{astro-ph/9710364}.

\bibitem[{\citenamefont{Kaplinghat et~al.}(2003)\citenamefont{Kaplinghat, Knox,
  and Song}}]{Kaplinghat:2003bh}
\bibinfo{author}{\bibfnamefont{M.}~\bibnamefont{Kaplinghat}},
  \bibinfo{author}{\bibfnamefont{L.}~\bibnamefont{Knox}}, \bibnamefont{and}
  \bibinfo{author}{\bibfnamefont{Y.-S.} \bibnamefont{Song}},
  \bibinfo{journal}{Phys. Rev. Lett.} \textbf{\bibinfo{volume}{91}},
  \bibinfo{pages}{241301} (\bibinfo{year}{2003}), \eprint{astro-ph/0303344}.

\bibitem[{\citenamefont{{Acquaviva} and
  {Baccigalupi}}(2006)}]{2006PhRvD..74j3510A}
\bibinfo{author}{\bibfnamefont{V.}~\bibnamefont{{Acquaviva}}} \bibnamefont{and}
  \bibinfo{author}{\bibfnamefont{C.}~\bibnamefont{{Baccigalupi}}},
  \bibinfo{journal}{Phys. Rev.} \textbf{\bibinfo{volume}{D74}},
  \bibinfo{eid}{103510} (\bibinfo{year}{2006}), \eprint{astro-ph/0507644}.

\bibitem[{\citenamefont{{Hu}}(2002)}]{2002PhRvD..65b3003H}
\bibinfo{author}{\bibfnamefont{W.}~\bibnamefont{{Hu}}}, \bibinfo{journal}{Phys.
  Rev.} \textbf{\bibinfo{volume}{D65}}, \bibinfo{eid}{023003}
  (\bibinfo{year}{2002}), \eprint{astro-ph/0108090}.

\bibitem[{\citenamefont{Lewis and Challinor}(2006)}]{Lewis:2006fu}
\bibinfo{author}{\bibfnamefont{A.}~\bibnamefont{Lewis}} \bibnamefont{and}
  \bibinfo{author}{\bibfnamefont{A.}~\bibnamefont{Challinor}},
  \bibinfo{journal}{Phys. Rept.} \textbf{\bibinfo{volume}{429}},
  \bibinfo{pages}{1} (\bibinfo{year}{2006}), \eprint{astro-ph/0601594}.

\bibitem[{\citenamefont{Smith et~al.}(2009)}]{2009AIPC.1141..121S}
\bibinfo{author}{\bibfnamefont{K.~M.} \bibnamefont{Smith}}
  \bibnamefont{et~al.}, \bibinfo{journal}{AIP Conf.Proc.}
  \textbf{\bibinfo{volume}{1141}}, \bibinfo{pages}{121} (\bibinfo{year}{2009}),
  \eprint{arXiv:0811.3916}.

\bibitem[{\citenamefont{Seljak}(1996)}]{Seljak:1995ve}
\bibinfo{author}{\bibfnamefont{U.}~\bibnamefont{Seljak}},
  \bibinfo{journal}{Astrophys. J.} \textbf{\bibinfo{volume}{463}},
  \bibinfo{pages}{1} (\bibinfo{year}{1996}), \eprint{astro-ph/9505109}.

\bibitem[{\citenamefont{Zaldarriaga and Seljak}(1998)}]{Zaldarriaga:1998ar}
\bibinfo{author}{\bibfnamefont{M.}~\bibnamefont{Zaldarriaga}} \bibnamefont{and}
  \bibinfo{author}{\bibfnamefont{U.}~\bibnamefont{Seljak}},
  \bibinfo{journal}{Phys. Rev.} \textbf{\bibinfo{volume}{D58}},
  \bibinfo{pages}{023003} (\bibinfo{year}{1998}), \eprint{astro-ph/9803150}.

\bibitem[{\citenamefont{Seljak and
  Zaldarriaga}(1999{\natexlab{a}})}]{Seljak:1998nu}
\bibinfo{author}{\bibfnamefont{U.}~\bibnamefont{Seljak}} \bibnamefont{and}
  \bibinfo{author}{\bibfnamefont{M.}~\bibnamefont{Zaldarriaga}},
  \bibinfo{journal}{Phys. Rev.} \textbf{\bibinfo{volume}{D60}},
  \bibinfo{pages}{043504} (\bibinfo{year}{1999}{\natexlab{a}}),
  \eprint{astro-ph/9811123}.

\bibitem[{\citenamefont{Zaldarriaga}(2000)}]{Zaldarriaga:2000ud}
\bibinfo{author}{\bibfnamefont{M.}~\bibnamefont{Zaldarriaga}},
  \bibinfo{journal}{Phys. Rev.} \textbf{\bibinfo{volume}{D62}},
  \bibinfo{pages}{063510} (\bibinfo{year}{2000}), \eprint{astro-ph/9910498}.

\bibitem[{\citenamefont{Seljak and
  Zaldarriaga}(1999{\natexlab{b}})}]{Seljak:1998aq}
\bibinfo{author}{\bibfnamefont{U.}~\bibnamefont{Seljak}} \bibnamefont{and}
  \bibinfo{author}{\bibfnamefont{M.}~\bibnamefont{Zaldarriaga}},
  \bibinfo{journal}{Phys. Rev. Lett.} \textbf{\bibinfo{volume}{82}},
  \bibinfo{pages}{2636} (\bibinfo{year}{1999}{\natexlab{b}}),
  \eprint{astro-ph/9810092}.

\bibitem[{\citenamefont{Hu and Okamoto}(2002)}]{Hu:2001kj}
\bibinfo{author}{\bibfnamefont{W.}~\bibnamefont{Hu}} \bibnamefont{and}
  \bibinfo{author}{\bibfnamefont{T.}~\bibnamefont{Okamoto}},
  \bibinfo{journal}{Astrophys. J.} \textbf{\bibinfo{volume}{574}},
  \bibinfo{pages}{566} (\bibinfo{year}{2002}), \eprint{astro-ph/0111606}.

\bibitem[{\citenamefont{Okamoto and Hu}(2003)}]{Okamoto:2003zw}
\bibinfo{author}{\bibfnamefont{T.}~\bibnamefont{Okamoto}} \bibnamefont{and}
  \bibinfo{author}{\bibfnamefont{W.}~\bibnamefont{Hu}}, \bibinfo{journal}{Phys.
  Rev.} \textbf{\bibinfo{volume}{D67}}, \bibinfo{pages}{083002}
  (\bibinfo{year}{2003}), \eprint{astro-ph/0301031}.

\bibitem[{\citenamefont{Hirata and Seljak}(2003{\natexlab{a}})}]{Hirata:2002jy}
\bibinfo{author}{\bibfnamefont{C.~M.} \bibnamefont{Hirata}} \bibnamefont{and}
  \bibinfo{author}{\bibfnamefont{U.}~\bibnamefont{Seljak}},
  \bibinfo{journal}{Phys. Rev.} \textbf{\bibinfo{volume}{D67}},
  \bibinfo{pages}{043001} (\bibinfo{year}{2003}{\natexlab{a}}),
  \eprint{astro-ph/0209489}.

\bibitem[{\citenamefont{Hirata and Seljak}(2003{\natexlab{b}})}]{Hirata:2003ka}
\bibinfo{author}{\bibfnamefont{C.~M.} \bibnamefont{Hirata}} \bibnamefont{and}
  \bibinfo{author}{\bibfnamefont{U.}~\bibnamefont{Seljak}},
  \bibinfo{journal}{Phys. Rev.} \textbf{\bibinfo{volume}{D68}},
  \bibinfo{pages}{083002} (\bibinfo{year}{2003}{\natexlab{b}}),
  \eprint{astro-ph/0306354}.

\bibitem[{\citenamefont{Challinor and Lewis}(2005)}]{Challinor:2005jy}
\bibinfo{author}{\bibfnamefont{A.}~\bibnamefont{Challinor}} \bibnamefont{and}
  \bibinfo{author}{\bibfnamefont{A.}~\bibnamefont{Lewis}},
  \bibinfo{journal}{Phys. Rev.} \textbf{\bibinfo{volume}{D71}},
  \bibinfo{pages}{103010} (\bibinfo{year}{2005}), \eprint{astro-ph/0502425}.

\bibitem[{\citenamefont{Cooray and Kesden}(2003)}]{Cooray:2002py}
\bibinfo{author}{\bibfnamefont{A.}~\bibnamefont{Cooray}} \bibnamefont{and}
  \bibinfo{author}{\bibfnamefont{M.}~\bibnamefont{Kesden}},
  \bibinfo{journal}{New Astron.} \textbf{\bibinfo{volume}{8}},
  \bibinfo{pages}{231} (\bibinfo{year}{2003}), \eprint{astro-ph/0204068}.

\bibitem[{\citenamefont{Kesden et~al.}(2003)\citenamefont{Kesden, Cooray, and
  Kamionkowski}}]{Kesden:2003cc}
\bibinfo{author}{\bibfnamefont{M.~H.} \bibnamefont{Kesden}},
  \bibinfo{author}{\bibfnamefont{A.}~\bibnamefont{Cooray}}, \bibnamefont{and}
  \bibinfo{author}{\bibfnamefont{M.}~\bibnamefont{Kamionkowski}},
  \bibinfo{journal}{Phys. Rev.} \textbf{\bibinfo{volume}{D67}},
  \bibinfo{pages}{123507} (\bibinfo{year}{2003}), \eprint{astro-ph/0302536}.

\bibitem[{\citenamefont{Hanson et~al.}(2011)\citenamefont{Hanson, Challinor,
  Efstathiou, and Bielewicz}}]{Hanson:2010rp}
\bibinfo{author}{\bibfnamefont{D.}~\bibnamefont{Hanson}},
  \bibinfo{author}{\bibfnamefont{A.}~\bibnamefont{Challinor}},
  \bibinfo{author}{\bibfnamefont{G.}~\bibnamefont{Efstathiou}},
  \bibnamefont{and}
  \bibinfo{author}{\bibfnamefont{P.}~\bibnamefont{Bielewicz}},
  \bibinfo{journal}{Phys. Rev.} \textbf{\bibinfo{volume}{D83}},
  \bibinfo{pages}{043005} (\bibinfo{year}{2011}), \eprint{arXiv:1008.4403}.

\bibitem[{\citenamefont{{Jenkins} et~al.}(2014)\citenamefont{{Jenkins},
  {Manohar}, {Waalewijn}, and {Yadav}}}]{2014arXiv1403.2386J}
\bibinfo{author}{\bibfnamefont{E.~E.} \bibnamefont{{Jenkins}}},
  \bibinfo{author}{\bibfnamefont{A.~V.} \bibnamefont{{Manohar}}},
  \bibinfo{author}{\bibfnamefont{W.~J.} \bibnamefont{{Waalewijn}}},
  \bibnamefont{and} \bibinfo{author}{\bibfnamefont{A.~P.~S.}
  \bibnamefont{{Yadav}}} (\bibinfo{year}{2014}), \eprint{arXiv:1403.2386}.

\bibitem[{\citenamefont{Anderes}(2013)}]{Anderes:2013jw}
\bibinfo{author}{\bibfnamefont{E.}~\bibnamefont{Anderes}}
  (\bibinfo{year}{2013}), \eprint{arXiv:1301.2576}.

\bibitem[{\citenamefont{Goroff et~al.}(1986)\citenamefont{Goroff, Grinstein,
  Rey, and Wise}}]{Goroff:1986ep}
\bibinfo{author}{\bibfnamefont{M.}~\bibnamefont{Goroff}},
  \bibinfo{author}{\bibfnamefont{B.}~\bibnamefont{Grinstein}},
  \bibinfo{author}{\bibfnamefont{S.}~\bibnamefont{Rey}}, \bibnamefont{and}
  \bibinfo{author}{\bibfnamefont{M.~B.} \bibnamefont{Wise}},
  \bibinfo{journal}{Astrophys. J.} \textbf{\bibinfo{volume}{311}},
  \bibinfo{pages}{6} (\bibinfo{year}{1986}).

\bibitem[{\citenamefont{Rathaus et~al.}(2011)\citenamefont{Rathaus, Fialkov,
  and Itzhaki}}]{Rathaus:2011xi}
\bibinfo{author}{\bibfnamefont{B.}~\bibnamefont{Rathaus}},
  \bibinfo{author}{\bibfnamefont{A.}~\bibnamefont{Fialkov}}, \bibnamefont{and}
  \bibinfo{author}{\bibfnamefont{N.}~\bibnamefont{Itzhaki}},
  \bibinfo{journal}{JCAP} \textbf{\bibinfo{volume}{1106}}, \bibinfo{pages}{033}
  (\bibinfo{year}{2011}), \eprint{arXiv:1105.2940}.

\bibitem[{\citenamefont{Bernardeau et~al.}(1997)\citenamefont{Bernardeau,
  Van~Waerbeke, and Mellier}}]{Bernardeau:1996un}
\bibinfo{author}{\bibfnamefont{F.}~\bibnamefont{Bernardeau}},
  \bibinfo{author}{\bibfnamefont{L.}~\bibnamefont{Van~Waerbeke}},
  \bibnamefont{and} \bibinfo{author}{\bibfnamefont{Y.}~\bibnamefont{Mellier}},
  \bibinfo{journal}{Astron. Astrophys.} \textbf{\bibinfo{volume}{322}},
  \bibinfo{pages}{1} (\bibinfo{year}{1997}), \eprint{astro-ph/9609122}.

\bibitem[{\citenamefont{Cooray and Hu}(2002)}]{Cooray:2002mj}
\bibinfo{author}{\bibfnamefont{A.}~\bibnamefont{Cooray}} \bibnamefont{and}
  \bibinfo{author}{\bibfnamefont{W.}~\bibnamefont{Hu}},
  \bibinfo{journal}{Astrophys. J.} \textbf{\bibinfo{volume}{574}},
  \bibinfo{pages}{19} (\bibinfo{year}{2002}), \eprint{astro-ph/0202411}.

\bibitem[{\citenamefont{Cooray et~al.}(2005)\citenamefont{Cooray, Kamionkowski,
  and Caldwell}}]{Cooray:2005hm}
\bibinfo{author}{\bibfnamefont{A.}~\bibnamefont{Cooray}},
  \bibinfo{author}{\bibfnamefont{M.}~\bibnamefont{Kamionkowski}},
  \bibnamefont{and} \bibinfo{author}{\bibfnamefont{R.~R.}
  \bibnamefont{Caldwell}}, \bibinfo{journal}{Phys. Rev.}
  \textbf{\bibinfo{volume}{D71}}, \bibinfo{pages}{123527}
  (\bibinfo{year}{2005}), \eprint{astro-ph/0503002}.

\bibitem[{\citenamefont{Namikawa et~al.}(2012)\citenamefont{Namikawa, Yamauchi,
  and Taruya}}]{Namikawa:2011cs}
\bibinfo{author}{\bibfnamefont{T.}~\bibnamefont{Namikawa}},
  \bibinfo{author}{\bibfnamefont{D.}~\bibnamefont{Yamauchi}}, \bibnamefont{and}
  \bibinfo{author}{\bibfnamefont{A.}~\bibnamefont{Taruya}},
  \bibinfo{journal}{JCAP} \textbf{\bibinfo{volume}{1201}}, \bibinfo{pages}{007}
  (\bibinfo{year}{2012}), \eprint{arXiv:1110.1718}.

\bibitem[{\citenamefont{Hu}(2000)}]{Hu:2000ee}
\bibinfo{author}{\bibfnamefont{W.}~\bibnamefont{Hu}}, \bibinfo{journal}{Phys.
  Rev.} \textbf{\bibinfo{volume}{D62}}, \bibinfo{pages}{043007}
  (\bibinfo{year}{2000}), \eprint{astro-ph/0001303}.

\bibitem[{\citenamefont{Yadav et~al.}(2010)\citenamefont{Yadav, Su, and
  Zaldarriaga}}]{Yadav:2009za}
\bibinfo{author}{\bibfnamefont{A.~P.} \bibnamefont{Yadav}},
  \bibinfo{author}{\bibfnamefont{M.}~\bibnamefont{Su}}, \bibnamefont{and}
  \bibinfo{author}{\bibfnamefont{M.}~\bibnamefont{Zaldarriaga}},
  \bibinfo{journal}{Phys. Rev.} \textbf{\bibinfo{volume}{D81}},
  \bibinfo{pages}{063512} (\bibinfo{year}{2010}), \eprint{arXiv:0912.3532}.

\bibitem[{\citenamefont{{Dvorkin} and {Smith}}(2009)}]{2009PhRvD..79d3003D}
\bibinfo{author}{\bibfnamefont{C.}~\bibnamefont{{Dvorkin}}} \bibnamefont{and}
  \bibinfo{author}{\bibfnamefont{K.~M.} \bibnamefont{{Smith}}},
  \bibinfo{journal}{Phys. Rev.} \textbf{\bibinfo{volume}{D79}},
  \bibinfo{eid}{043003} (\bibinfo{year}{2009}), \eprint{arXiv:0812.1566}.

\bibitem[{\citenamefont{{Yadav} et~al.}(2009)\citenamefont{{Yadav}, {Biswas},
  {Su}, and {Zaldarriaga}}}]{2009PhRvD..79l3009Y}
\bibinfo{author}{\bibfnamefont{A.~P.~S.} \bibnamefont{{Yadav}}},
  \bibinfo{author}{\bibfnamefont{R.}~\bibnamefont{{Biswas}}},
  \bibinfo{author}{\bibfnamefont{M.}~\bibnamefont{{Su}}}, \bibnamefont{and}
  \bibinfo{author}{\bibfnamefont{M.}~\bibnamefont{{Zaldarriaga}}},
  \bibinfo{journal}{Phys. Rev.} \textbf{\bibinfo{volume}{D79}},
  \bibinfo{eid}{123009} (\bibinfo{year}{2009}), \eprint{arXiv:0902.4466}.

\bibitem[{\citenamefont{{Gluscevic} et~al.}(2009)\citenamefont{{Gluscevic},
  {Kamionkowski}, and {Cooray}}}]{2009PhRvD..80b3510G}
\bibinfo{author}{\bibfnamefont{V.}~\bibnamefont{{Gluscevic}}},
  \bibinfo{author}{\bibfnamefont{M.}~\bibnamefont{{Kamionkowski}}},
  \bibnamefont{and} \bibinfo{author}{\bibfnamefont{A.}~\bibnamefont{{Cooray}}},
  \bibinfo{journal}{Phys. Rev.} \textbf{\bibinfo{volume}{D80}},
  \bibinfo{eid}{023510} (\bibinfo{year}{2009}), \eprint{arXiv:0905.1687}.

\bibitem[{\citenamefont{Kamionkowski}(2009)}]{2009PhRvL.102k1302K}
\bibinfo{author}{\bibfnamefont{M.}~\bibnamefont{Kamionkowski}},
  \bibinfo{journal}{Phys. Rev. Lett.} \textbf{\bibinfo{volume}{102}},
  \bibinfo{pages}{111302} (\bibinfo{year}{2009}), \eprint{arXiv:0810.1286}.

\bibitem[{\citenamefont{{Yadav} et~al.}(2012)\citenamefont{{Yadav}, {Pogosian},
  and {Vachaspati}}}]{2012PhRvD..86l3009Y}
\bibinfo{author}{\bibfnamefont{A.}~\bibnamefont{{Yadav}}},
  \bibinfo{author}{\bibfnamefont{L.}~\bibnamefont{{Pogosian}}},
  \bibnamefont{and}
  \bibinfo{author}{\bibfnamefont{T.}~\bibnamefont{{Vachaspati}}},
  \bibinfo{journal}{Phys. Rev.} \textbf{\bibinfo{volume}{D86}},
  \bibinfo{eid}{123009} (\bibinfo{year}{2012}), \eprint{arXiv:1207.3356}.

\bibitem[{\citenamefont{{Goldberg} and {Spergel}}(1999)}]{1999PhRvD..59j3002G}
\bibinfo{author}{\bibfnamefont{D.~M.} \bibnamefont{{Goldberg}}}
  \bibnamefont{and} \bibinfo{author}{\bibfnamefont{D.~N.}
  \bibnamefont{{Spergel}}}, \bibinfo{journal}{Phys. Rev.}
  \textbf{\bibinfo{volume}{D59}}, \bibinfo{eid}{103002} (\bibinfo{year}{1999}),
  \eprint{astro-ph/9811251}.

\bibitem[{\citenamefont{{Spergel} and {Goldberg}}(1999)}]{1999PhRvD..59j3001S}
\bibinfo{author}{\bibfnamefont{D.~N.} \bibnamefont{{Spergel}}}
  \bibnamefont{and} \bibinfo{author}{\bibfnamefont{D.~M.}
  \bibnamefont{{Goldberg}}}, \bibinfo{journal}{Phys. Rev.}
  \textbf{\bibinfo{volume}{D59}}, \bibinfo{eid}{103001} (\bibinfo{year}{1999}),
  \eprint{astro-ph/9811252}.

\bibitem[{\citenamefont{{Zaldarriaga} and
  {Seljak}}(1999)}]{1999PhRvD..59l3507Z}
\bibinfo{author}{\bibfnamefont{M.}~\bibnamefont{{Zaldarriaga}}}
  \bibnamefont{and} \bibinfo{author}{\bibfnamefont{U.}~\bibnamefont{{Seljak}}},
  \bibinfo{journal}{Phys. Rev.} \textbf{\bibinfo{volume}{D59}},
  \bibinfo{eid}{123507} (\bibinfo{year}{1999}), \eprint{astro-ph/9810257}.

\bibitem[{\citenamefont{Sachs and Wolfe}(1967)}]{ISW}
\bibinfo{author}{\bibfnamefont{R.~K.} \bibnamefont{Sachs}} \bibnamefont{and}
  \bibinfo{author}{\bibfnamefont{A.~M.} \bibnamefont{Wolfe}},
  \bibinfo{journal}{Astrophys. J.} \textbf{\bibinfo{volume}{147}},
  \bibinfo{pages}{73} (\bibinfo{year}{1967}).

\bibitem[{\citenamefont{Sunyaev and Zeldovich}(1980)}]{SZ}
\bibinfo{author}{\bibfnamefont{R.~A.} \bibnamefont{Sunyaev}} \bibnamefont{and}
  \bibinfo{author}{\bibfnamefont{Y.~B.} \bibnamefont{Zeldovich}},
  \bibinfo{journal}{MNRAS} \textbf{\bibinfo{volume}{190}}, \bibinfo{pages}{413}
  (\bibinfo{year}{1980}).

\bibitem[{\citenamefont{Brown}(1992)}]{Brown:1992db}
\bibinfo{author}{\bibfnamefont{L.}~\bibnamefont{Brown}},
  \emph{\bibinfo{title}{{Quantum Field Theory}}} (\bibinfo{year}{1992}).

\bibitem[{\citenamefont{Peskin and Schroeder}(1995)}]{Peskin:1995ev}
\bibinfo{author}{\bibfnamefont{M.~E.} \bibnamefont{Peskin}} \bibnamefont{and}
  \bibinfo{author}{\bibfnamefont{D.~V.} \bibnamefont{Schroeder}},
  \emph{\bibinfo{title}{{An Introduction to Quantum Field Theory}}}
  (\bibinfo{year}{1995}).

\bibitem[{\citenamefont{{Cooray} and {Hu}}(2000)}]{2000ApJ...534..533C}
\bibinfo{author}{\bibfnamefont{A.}~\bibnamefont{{Cooray}}} \bibnamefont{and}
  \bibinfo{author}{\bibfnamefont{W.}~\bibnamefont{{Hu}}},
  \bibinfo{journal}{\apj} \textbf{\bibinfo{volume}{534}}, \bibinfo{pages}{533}
  (\bibinfo{year}{2000}), \eprint{astro-ph/9910397}.

\bibitem[{\citenamefont{Kosowsky and Loeb}(1996)}]{Kosowsky:1996yc}
\bibinfo{author}{\bibfnamefont{A.}~\bibnamefont{Kosowsky}} \bibnamefont{and}
  \bibinfo{author}{\bibfnamefont{A.}~\bibnamefont{Loeb}},
  \bibinfo{journal}{Astrophys. J.} \textbf{\bibinfo{volume}{469}},
  \bibinfo{pages}{1} (\bibinfo{year}{1996}), \eprint{astro-ph/9601055}.

\bibitem[{\citenamefont{{Kosowsky} et~al.}(2005)\citenamefont{{Kosowsky},
  {Kahniashvili}, {Lavrelashvili}, and {Ratra}}}]{2005PhRvD..71d3006K}
\bibinfo{author}{\bibfnamefont{A.}~\bibnamefont{{Kosowsky}}},
  \bibinfo{author}{\bibfnamefont{T.}~\bibnamefont{{Kahniashvili}}},
  \bibinfo{author}{\bibfnamefont{G.}~\bibnamefont{{Lavrelashvili}}},
  \bibnamefont{and} \bibinfo{author}{\bibfnamefont{B.}~\bibnamefont{{Ratra}}},
  \bibinfo{journal}{Phys. Rev.} \textbf{\bibinfo{volume}{D71}},
  \bibinfo{pages}{043006} (\bibinfo{year}{2005}).

\bibitem[{\citenamefont{{Hu} et~al.}(2003)\citenamefont{{Hu}, {Hedman}, and
  {Zaldarriaga}}}]{2003PhRvD..67d3004H}
\bibinfo{author}{\bibfnamefont{W.}~\bibnamefont{{Hu}}},
  \bibinfo{author}{\bibfnamefont{M.~M.} \bibnamefont{{Hedman}}},
  \bibnamefont{and}
  \bibinfo{author}{\bibfnamefont{M.}~\bibnamefont{{Zaldarriaga}}},
  \bibinfo{journal}{Phys. Rev.} \textbf{\bibinfo{volume}{D67}},
  \bibinfo{eid}{043004} (\bibinfo{year}{2003}), \eprint{astro-ph/0210096}.

\bibitem[{\citenamefont{{Carroll}}(1998)}]{1998PhRvL..81.3067C}
\bibinfo{author}{\bibfnamefont{S.~M.} \bibnamefont{{Carroll}}},
  \bibinfo{journal}{Phys. Rev. Lett.} \textbf{\bibinfo{volume}{81}},
  \bibinfo{pages}{3067} (\bibinfo{year}{1998}), \eprint{astro-ph/9806099}.

\bibitem[{\citenamefont{{Pospelov} et~al.}(2009)\citenamefont{{Pospelov},
  {Ritz}, and {Skordis}}}]{2009PhRvL.103e1302P}
\bibinfo{author}{\bibfnamefont{M.}~\bibnamefont{{Pospelov}}},
  \bibinfo{author}{\bibfnamefont{A.}~\bibnamefont{{Ritz}}}, \bibnamefont{and}
  \bibinfo{author}{\bibfnamefont{C.}~\bibnamefont{{Skordis}}},
  \bibinfo{journal}{Phys. Rev. Lett.} \textbf{\bibinfo{volume}{103}},
  \bibinfo{eid}{051302} (\bibinfo{year}{2009}), \eprint{arXiv:0808.0673}.

\bibitem[{\citenamefont{Fan et~al.}(2006)\citenamefont{Fan, Strauss, Becker,
  White, Gunn et~al.}}]{2006AJ....132..117F}
\bibinfo{author}{\bibfnamefont{X.-H.} \bibnamefont{Fan}},
  \bibinfo{author}{\bibfnamefont{M.~A.} \bibnamefont{Strauss}},
  \bibinfo{author}{\bibfnamefont{R.~H.} \bibnamefont{Becker}},
  \bibinfo{author}{\bibfnamefont{R.~L.} \bibnamefont{White}},
  \bibinfo{author}{\bibfnamefont{J.~E.} \bibnamefont{Gunn}},
  \bibnamefont{et~al.}, \bibinfo{journal}{Astron. J.}
  \textbf{\bibinfo{volume}{132}}, \bibinfo{pages}{117} (\bibinfo{year}{2006}),
  \eprint{astro-ph/0512082}.

\bibitem[{\citenamefont{Hinshaw et~al.}(2013)}]{2013ApJS..208...19H}
\bibinfo{author}{\bibfnamefont{G.}~\bibnamefont{Hinshaw}} \bibnamefont{et~al.}
  (\bibinfo{collaboration}{WMAP}), \bibinfo{journal}{Astrophys. J. Suppl.}
  \textbf{\bibinfo{volume}{208}}, \bibinfo{pages}{19} (\bibinfo{year}{2013}),
  \eprint{arXiv:1212.5226}.

\bibitem[{\citenamefont{Knox}(1995)}]{Knox:1995dq}
\bibinfo{author}{\bibfnamefont{L.}~\bibnamefont{Knox}}, \bibinfo{journal}{Phys.
  Rev.} \textbf{\bibinfo{volume}{D52}}, \bibinfo{pages}{4307}
  (\bibinfo{year}{1995}), \eprint{astro-ph/9504054}.

\end{thebibliography}

\end{document}